\documentclass[10pt,aps,prb,twocolumn,noshowpacs]{revtex4-1}
\usepackage{amsmath,graphicx,amsbsy,amssymb,amsmath,epsfig,latexsym,wasysym,pifont,mathrsfs,natbib}
\usepackage{float} \newfloat{widefig}{thp}{lop} \usepackage{comment}

\usepackage{array, makecell} \usepackage{verbatim} \usepackage{datetime}
\usepackage{multirow,array} \usepackage{hyperref} \usepackage{color} \usepackage{setspace}
\usepackage{subcaption}
\usepackage{graphicx}
\usepackage[export]{adjustbox}

\hypersetup{colorlinks,%,%
  linkcolor=blue,%
  citecolor=blue,%
  urlcolor=blue}

\begin{document}
\title{\textbf{Stacking and Layer dependence of magnetic properties in Ti\textsubscript{2}C and Fe\textsubscript{2}C}}

\author{Himangshu Sekhar Sarmah}
\email[]{shimangshu@iitg.ac.in}
\affiliation{Department of Physics, Indian Institute of Technology
  Guwahati, Guwahati-781039, Assam, India.}    
\author{Subhradip Ghosh}
\email{subhra@iitg.ac.in} \affiliation{Department of Physics,
  Indian Institute of Technology Guwahati, Guwahati-781039, Assam,
  India.} 
  \begin{abstract}  
Magnetic MXenes are turning out to be an important family of materials for exploring 2D magnetism. However, investigations into the inter-dependence of layer thickness, stacking patterns and magnetism in these materials, from a microscopic point of view, is still lacking. In this work, we have used Density Functional Theory (DFT) based calculations to understand the effects of layer thickness and stacking on the magnetic properties in two magnetic MXenes, Ti$_{2}$C and Fe$_{2}$C in their monolayer and bilayer forms. The ground state magnetic structures, magnetic moments, magnetic exchange interactions, magnetic transition temperatures and magnetic anisotropy energies are calculated and analysed using their electronic structures and standardised models. We find that in both systems increase in layer thickness (monolayer to bilayer) affects the ground state magnetic configuration which is driven by the changes in the magnetic exchange interactions. While the effects of stacking pattern is rather weak in Ti$_{2}$C, they are substantial, both qualitatively and quantitatively in Fe$_{2}$C. The computed results are analysed from their electronic structures. The results suggest that fascinating physical effects can be obtained in Fe$_{2}$C by tuning the layer thickness and stacking patterns, making it more suitable for device applications.
\end{abstract}


\maketitle

\section{Introduction\label{intro}}
Since the discovery of single-layer graphene in 2004 \cite{novoselov2004electric}, two-dimensional (2D) materials have attracted huge interest. In less than two decades huge family of 2D materials like transition metal dichalcogenides (TMDs)\cite{chhowalla2015two},hexagonal Boron Nitride\cite{pacile2008two},Silicene\cite{vogt2012silicene},phosphorene \cite{kou2015phosphorene} are discovered. These 2D materials, due to their reduced dimensions, exhibit many novel physical properties not seen in their bulk counterparts. In the age of miniaturisation, they have made significant inroads into various areas of applications like electronics \cite{roy2014field,kajale20212d,zhang2018two}, and photonics\cite{subbanna20222d}. The easy integrability of 2D materials to form heterostructures is an added advantage since it provides freedom and flexibility to devise new materials with functional properties. 

Magnetism in  2D materials has been a subject of contemporary interest. Attempts to alleviate the  skepticism regarding discovery of intrinsic magnetism in 2D materials as established by the Mermin Wagner theorem \cite{mermin} was addressed by considering the possibility of magnetic anisotropy stabilzing magnetism in 2D. This idea gained currency once magnetic ordering in CrI\textsubscript{3} was observed in 2017\cite{huang2017layer}. The experiment found intrinsic ferromagnetic ordering in CrI\textsubscript{3} with Curie temperature of around 45 K.After that, various 2D magnetic materials like VSe\textsubscript{2}\cite{wang2021ferromagnetism},Cr\textsubscript{2}Ge\textsubscript{2}Te\textsubscript{6}\cite{gong2017discovery},Fe\textsubscript{3}GeTe\textsubscript{2}\cite{fei2018two} etc were successfully synthesised and studied.The existence of the long-range ordering in these 2D materials are attributed to their large magnetic anisotropy. The 2D materials, due to their flexibility, high sensitivity, and susceptibility to external perturbations, hold promises regarding applications in spintronics and quantum technologies. 

Little more than a decade ago a new family of 2D materials called MXene was discovered upon exfoliation of  layered 3D MAX compounds \cite{naguib2011two}. The exfoliation to MXene from MAX are done by removing the A layers that consist of elements from group 13 or 14.The resulting 2D MXenes (M a transition metal and X either C or N) have chemical composition  M$_{n+1}$X$_{n}$  where $n = 1-3$. Since the first discovery in 2011, quite a few  MXenes have been syntheseised \cite{naguib2012two,naguib2013new,naguib2013new,soundiraraju2017two,meshkian2015synthesis,urbankowski20172d,urbankowski2016synthesis,naguib2011two,zhou2016two}. The major area of application of the MXenes so far is in   
batteries and supercapacitors\cite{ma2021ti,zheng2022mxene}. Comparatively, exploration of magnetism in MXenes is substantially less although the compounds offer more tunability in terms of constituents and compositions. Majority of the work on magnetism in MXenes are theoretical, mostly through first principles electronic structure calculations. However limited the exploration is, the outcome has been quite exciting as both ferromagnetic  and antiferromagnetic ordering have been found in them \cite{si2015half,sun2021cr2nx2,yue2021tuning,he2016new,zhang2022computational}. .

Though these calculations provide important aid in understanding magnetism in MXenes, one major limitation is that the works are done on monolayers while in reality experimentally synthesised MXenes are multilayered in general. The magnetic properties of 2D materials are found to be dependent crucially on the thickness of layers and their stacking patterns \cite{huang2017layer,sivadas2018stacking,science2019,mxene-multilayer2019,kong2021switching,wang2021systematic} due to the variations in intra and inter-layer exchange interactions. With the help of Density functional theory (DFT) based first - principles calculations, in this work, we investigate the layer and stacking pattern dependencies of magnetic properties in Ti\textsubscript{2}C, and Fe\textsubscript{2}C MXenes. These two systems have different magnetic ground states in monolayers. While Ti\textsubscript{2}C has AFM long range order as the monolayer ground state, Fe\textsubscript{2}C monolayer has FM ordering in its ground state\cite{lv2020monolayer,yue2017fe2c}. Such differences in the monolayer ground states motivated us to consider these two systems for investigations into the multilayer effects on magnetic properties in MXenes. For this purpose, we have considered monolayers and bi-layers only and looked into the changes in the electronic structures, magnetic exchange interactions, magnetic transition temperatures and magnetic anisotropy energies with variations in layer thickness and stacking. Our work, for the first time, addressed the aspect of multilayer-magnetism relation in MXenes. Our results illustrate the differences in the nature of magnetism in these two systems and provides important insights which can be useful for using these compounds in device applications
\section{Computational Details}
\begin{figure*}
    \includegraphics[height=8cm, width=15.00 cm]{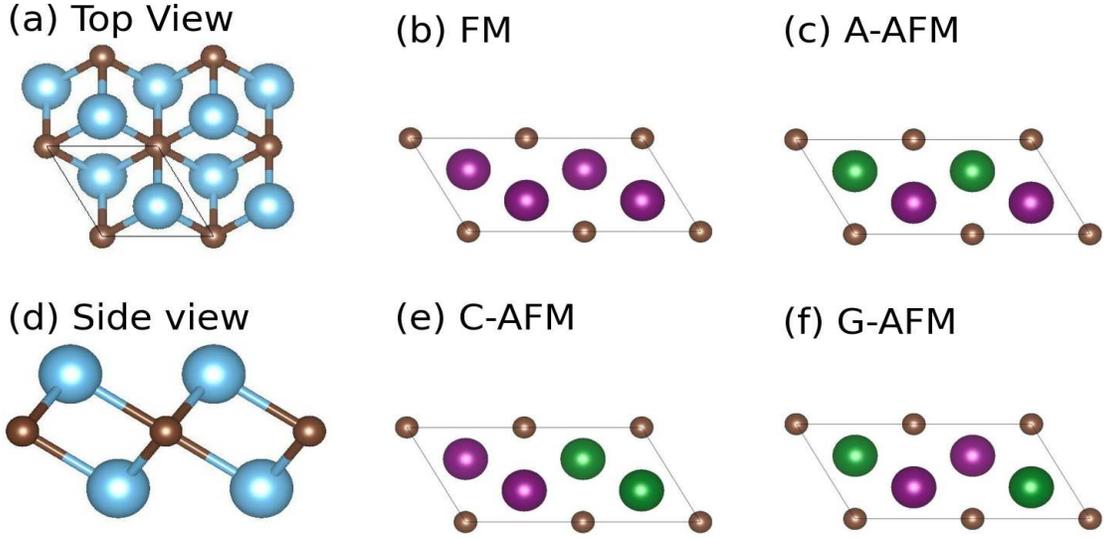}
 \caption{(a) and (d) are top and side views of M\textsubscript{2}C monolayer; M and C atoms are represented by blue and brown balls respectively.(b),(c),(e),f) are the four possible magnetic configurations in monolayer M\textsubscript{2}C.Purple and green colour stand for spin up and spin down configurations respectively.}
    \label{Fig:1} 
\end{figure*}
All calculations are done within the framework of density functional theory (DFT)\cite{dft} using the plane wave basis set and the  projector augmented wave (PAW) pseudopotentials\cite{kresse1999ultrasoft} as implemented in the Vienna Ab initio Simulation Package (VASP)\cite{kresse1996efficient}. The exchange-correlation part of the Hamiltonian is approximated using generalized gradient approximation (GGA) as parameterized by Perdew-Burke-Ernzerhof (PBE) \cite{perdew1996generalized}. The plane wave cut-off energy is set as 600 eV after carefully checking its convergence. The Van Der Waals interactions are addressed using the DFT-D3 \cite{grimme2010consistent}. The convergence criteria is set as $ 10^{-6} $ for energy and 0.001 eV/{\AA} for force. $ 12\times 12\times 1 $ and $ 36\times 36\times 1 $ dense Monkhorst pack \cite{monkhorst1976special} $k$-meshes have been used for structural relaxations and electronic structure calculations, respectively. Spin-orbit coupling (SOC) has also taken into account for calculations of magneto-crystalline anisotropy energies (MAE). The magnetic exchange interactions are calculated by mapping DFT total energies on to a 2D Ising Hamiltonian. These are then used to calculate the magnetic transition temperatures by classical Monte Carlo simulations with a Heisenberg model as implemented in {\it UppASD} code \cite{eriksson2017atomistic}. 

The structure of monolayer M\textsubscript{2}C (M=Ti,Fe) is shown in Figure ~\ref{Fig:1}(a),(d). Monolayer M\textsubscript{2}C, consists of two M layers at the top and a C-layer sandwiched between them. A vacuum of 20{\AA} is considered in the normal direction to avoid interactions between the periodic images. All calculations are done on $ 2\times 1\times 1 $  supercells since it was necessary to represent bi-layers as well as various magnetic configurations. 

\section{Results and discussion}
\begin{figure*}
    \includegraphics[height=8cm, width=15.00 cm]{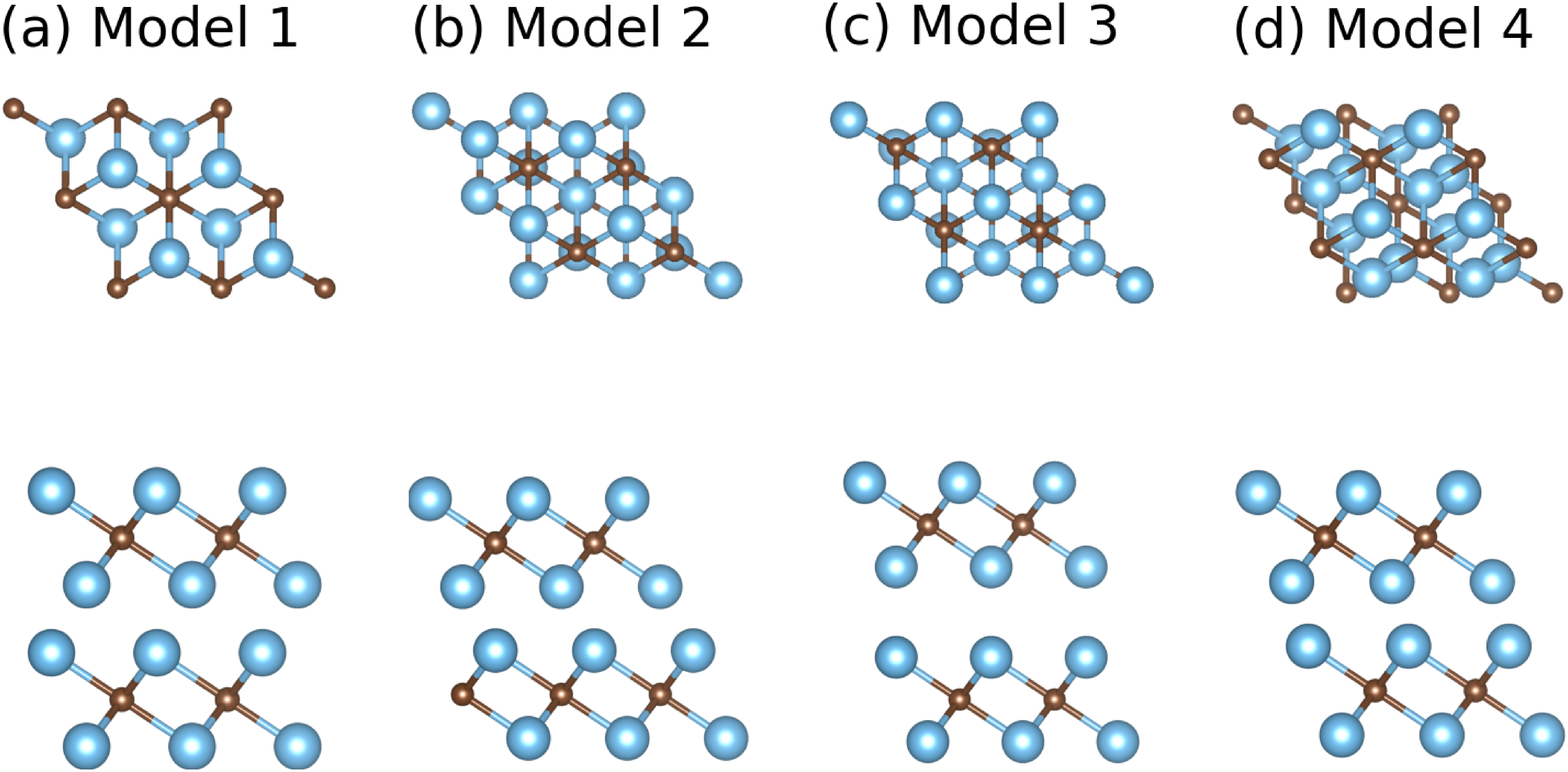}
     \caption{Four stacking configurations of bilayer Mi\textsubscript{2}C.Both top view (upper row ) and side view ( lower row ) are included for clarity.The blue atoms indicates the M(M=Ti,Fe) atom where as the brown ones are the C atoms}
    \label{Fig:2}   
\end{figure*}
\subsection{ Structural and magnetic ground states}
The ground states of monolayer and bi-layer systems are obtained by optimising a non-magnetic structure followed by another optimisation over various magnetic configurations. 
The lattice parameters of Ti\textsubscript{2}C and Fe\textsubscript{2}C are 3.01{\AA} and 2.62{\AA} respectively, in good agreement with existing results\cite{lv2020monolayer}. To optimise the non-magnetic bilayer structure of Ti\textsubscript{2}C and Fe\textsubscript{2}C we consider four different possible stacking configurations (Model 1-4)\cite{huang2021theoretical}.These are shown in 
Figure ~\ref{Fig:2}. The stacking patterns differ with each other the following way: in Model 1, the C atom of the top layer is directly above the C atom of the bottom layer.In Model 2 (Model 3),the C atom of top layer is directly above the upper(lower) Ti atom of the bottom layer.In Model 4,the C atom of top layer is located directly above a hollow site in between the two Ti atoms  of the bottom layer.
 Upon relaxation, we find that with the stacking pattern in Model 4 the structure undergoes significant distortion for both MXenes. The optimized inter-layer Ti-Ti (Fe-Fe) separations for Ti\textsubscript{2}C(Fe\textsubscript{2}C) are 3.04(2.61) {\AA}, 3.02(2.62) {\AA} and 2.79(2.31) {\AA} in cases of Model 1, Model 2 and Model 3 respectively. Since compounds with stacking patterns according to Model 4 distorts the structure and  Model 3 of Ti\textsubscript{2}C is reported to be dynamically unstable \cite{huang2021theoretical}, we discard these structures at this stage and work with the rest to find ground state magnetic configurations. \\
\begin{table}
\caption{\label{tab:Table1} Equilibrium lattice constants and total energies of different magnetic configurations of monolayer M\textsubscript{2}C MXenes. Energy zero is set for the configurations with minimum total energy. }
\begin{tabular}{m{0.07\textwidth}m{0.15\textwidth}m{0.12\textwidth}m{0.1\textwidth}}
 \hline
Element & Magnetic Configurations & lattice-paranmeter( {\AA})& Energy(meV) \\ 
 \hline
 Ti\textsubscript{2}C & NM & 3.01 & 237.36 \\ 
  & FM & 3.05 & 73.68 \\ 
   & A-AFM & 3.04 & 0 \\ 
  & C-AFM & 3.04 & 206.55 \\ 
   & G-AFM & 3.04 & 198.06 \\ 
 \hline
 Fe\textsubscript{2}C & NM & 2.62 & 1532.34 \\ 
   & FM & 2.81 & 0 \\ 
 & A-AFM & 2.82 & 295.47 \\ 
  & C-AFM & 2.80 & 424.04 \\ 
  & G-AFM & 2.81 & 466.49 \\ 
 \hline
 \end{tabular}
 \end{table}
 \begin{table*}
\caption{\label{tab:Table2}Total energies of various magnetic configurations considered for bilayer Ti\textsubscript{2}C. The indexes 1-8 stand for the Ti atoms in bilayer. $\uparrow$($\downarrow$) denotes the alignment of spin along (opposite to) $z$-direction. The lowest energy configuration is set as Energy zero. Energies are in units of meV.}
\begin{tabular}{ m{0.08\textwidth}|m{0.06\textwidth}m{0.06\textwidth}m{0.06\textwidth}m{0.06\textwidth}m{0.06\textwidth}m{0.06\textwidth}m{0.06\textwidth}m{0.06\textwidth}m{0.15\textwidth}m{0.15\textwidth}}
 \hline
 Type & 1 & 2 & 3 & 4 & 5 & 6 & 7 & 8 & Ti\textsubscript{2}C-Model 1  & Ti\textsubscript{2}C-Model2 \\ 
 \hline
 FM & $ \uparrow $  & $ \uparrow $  & $ \uparrow $  & $ \uparrow $  & $ \uparrow $  & $ \uparrow $  & $ \uparrow $  & $ \uparrow $  & 0 & 0 \\ 
 
 AFM1 & $ \uparrow $  & $ \uparrow $  & $ \downarrow $ & $ \downarrow $ & $ \uparrow $  & $ \uparrow $  & $ \downarrow $ & $ \downarrow $ & 12.04 & 19.13 \\ 
 
 AFM2 & $ \uparrow $  & $ \downarrow $ & $ \uparrow $  & $ \downarrow $ & $ \uparrow $  & $ \downarrow $ & $ \uparrow $  & $ \downarrow $ & 176.87 & 176.78 \\
 
 AFM3 & $ \downarrow $ & $ \uparrow $  & $ \uparrow $  & $ \downarrow $ & $ \downarrow $ & $ \uparrow $  & $ \uparrow $  & $ \downarrow $ & 170.80 & 153.61\\
 
 AFM4  & $ \uparrow $  & $ \uparrow $  & $ \uparrow $  & $ \uparrow $  & $ \downarrow $ & $ \downarrow $ & $ \downarrow $ & $ \downarrow $ & 12.02  & 15.32\\
 
 AFM5 & $ \uparrow $  & $ \uparrow $  & $ \downarrow $ & $ \downarrow $ & $ \uparrow $  & $ \downarrow $ & $ \uparrow $  & $ \downarrow $ & 2.59 & 15.32\\
 
 AFM6 & $ \uparrow $  & $ \uparrow $  & $ \downarrow $ & $ \downarrow $ & $ \downarrow $ & $ \uparrow $  & $ \uparrow $  & $ \downarrow $ &  2.59 & 15.31\\
 
 AFM7 & $ \uparrow $  & $ \downarrow $ & $ \uparrow $  & $ \downarrow $ & $ \downarrow $ & $ \uparrow $  & $ \uparrow $  & $ \downarrow $ & 166.32 & 153.63\\
 \hline
\end{tabular}
\end{table*}
\begin{table*}
\caption{\label{tab:Table3}Total energies of various magnetic configurations considered for bilayer Fe\textsubscript{2}C. The indexes 1-8 stand for the Fe atoms in bilayer. $\uparrow$($\downarrow$) denotes the alignment of spin along (opposite to) $z$-direction. The lowest energy configuration is set as Energy zero. Energies are in units of meV.}
\begin{tabular}{ m{0.06\textwidth}|m{0.04\textwidth}m{0.04\textwidth}m{0.06\textwidth}m{0.04\textwidth}m{0.04\textwidth}m{0.04\textwidth}m{0.04\textwidth}m{0.04\textwidth}m{0.15\textwidth}m{0.15\textwidth}m{0.15\textwidth}}
 \hline
 Type & 1 & 2 & 3 & 4 & 5 & 6 & 7 & 8 & Fe\textsubscript{2}C-Model 1  & Fe\textsubscript{2}C-Model 2 & Fe\textsubscript{2}C-Model 3\\ 
 \hline
 FM & $ \uparrow $ & $ \uparrow $ & $ \uparrow $ & $ \uparrow $ & $ \uparrow $ & $ \uparrow $ & $ \uparrow $ & $ \uparrow $ & 0 & 0 & 62.23 \\ 
 
 AFM1 & $ \uparrow $ & $ \uparrow $ & $ \downarrow $ & $ \downarrow $ & $ \uparrow $ & $ \uparrow $ & $ \downarrow $ & $ \downarrow $ & 1164.23 & 825.32 & deformed  \\ 
 
 AFM2 & $ \uparrow $ & $ \downarrow $ & $ \uparrow $ & $ \downarrow $ & $ \uparrow $ & $ \downarrow $ & $ \uparrow $ & $ \downarrow $ & 1105.57 & Deformed  & deformed\\
 
 AFM3 & $ \downarrow $ & $ \uparrow $ & $ \uparrow $ & $ \downarrow $ & $ \downarrow $ & $ \uparrow $ & $ \uparrow $ & $ \downarrow $ & 1250.59 & 650.65 & 766.72\\
 AFM4  & $ \uparrow $ & $ \uparrow $ & $ \uparrow $ & $ \uparrow $ & $ \downarrow $ & $ \downarrow $ & $ \downarrow $ & $ \downarrow $ & 481.22  & 381.04 & 0 \\
 AFM5 & $ \uparrow $ & $ \uparrow $ & $ \downarrow $ & $ \downarrow $ & $ \uparrow $ & $ \downarrow $ & $ \uparrow $ & $ \downarrow $ & 951.28 & 386.40 & deformed \\ 
 AFM6 & $ \uparrow $ & $ \uparrow $ & $ \downarrow $ & $ \downarrow $ & $ \downarrow $ & $ \uparrow $ & $ \uparrow $ & $ \downarrow $ &  deformed & 386.41 & deformed\\
 
 AFM7 & $ \uparrow $ & $ \downarrow $ & $ \uparrow $ & $ \downarrow $ & $ \downarrow $ & $ \uparrow $ & $ \uparrow $ & $ \downarrow $ & Deformed & Deformed & 797.22\\
 \hline
\end{tabular}
\end{table*}
%\subsection{Magnetic Groundstate}

To find the magnetic ground states of monolayer Ti\textsubscript{2}C and Fe\textsubscript{2}C, four differrent magnetic configurations are considered (Figure ~\ref{Fig:1}). The calculated total energies of these configurations along with their lattice parameters are shown in the Table \ref{tab:Table1}. We find that for Ti\textsubscript{2}C, A-AFM configuration has the lowest energy. The lattice constant of Ti\textsubscript{2}C in A-AFM is 3.04 {\AA}. The individual magnetic moment of Ti atoms is 0.56 $ \mu B$. This result is in very good agreement with 
Reference~\onlinecite{lv2020monolayer} . For Fe\textsubscript{2}C, FM or ferromagnetic structure is the most stable structure with a high Fe magnetic moment of 1.95 $ \mu B $. This result, too, is in very good agreement with Reference~ \onlinecite{yue2017fe2c}. 

 To determine the magnetic ground state of bilayers, we considered eight different magnetic configurations. These eight possible configurations are summarised in Tables ~\ref{tab:Table2}, ~\ref{tab:Table3} and Figure S1, supplementary information. Indexed 1-4 are the M atoms in the bottom layers of M$_{2}$C MXene while ones indexed 5-8 are the M atoms in the top layer. The energies of all magnetic configurations for each one of the stacking models are compared with respect to the one having minimum energy (with the same stacking pattern).
 
 Ti\textsubscript{2}C bilayer in both models (Model 1 and 2) has ferromagnet (FM) configuration as its ground state. However, the magnetic moments of all the Ti atoms are not equal. The inner surface Ti atoms of both layers have a near vanishing magnetic moment of 0.04(-0.05) $ \mu B $ in Model 1(Model 2). The outer surface Ti atoms of both layers have a magnetic moment of 0.55 (0.47) $ \mu B $ in Model 1( Model 2). Thus the magnetism in bilayer Ti$_{2}$C is  controlled by the Ti atoms on the outer surfaces. 
This is a significant departure from what is observed in case of monolayer. In monolayer Ti\textsubscript{2}C, the chemical environment surrounding each Ti  atom is the same. Hence the magnetic moments of all Ti atoms are same. However, in the case of a bilayer, the chemical environments of the inner and outer surface Ti atoms are different, resulting in  unequal magnetic moments on them. The near vanishing magnetic moment of inner Ti atoms implies that the magnetism originates from the dangling states of the outer surface Ti atoms. Such surface dangling bond driven magnetism has been observed in low dimensional materials like Gallium nitride nanoclusters \cite{GaN}. 

 Fe$_{2}$C with stacking patterns of Model 1 and Model 2 retain its FM ground state in bi-layer structure.
 Similar behavior like  bilayer Ti\textsubscript{2}C with regard to atomic magnetic moment is observed in case of this MXene. But unlike Ti\textsubscript{2}C, all Fe atoms have significant magnetic moments in bilayer Fe$_{2}$C and they vary substantially as stacking pattern changes. The magnetic moment of inner surface Fe is 1.93 (1.51)$ \mu_B$ in Model 1 (Model 2). The outer surface Fe has a moment of 1.77 $\mu_{B}$ in Model 1 and 1.81 $\mu_{B}$ in Model 2. This implies that the stacking pattern affects the inner surface Fe atoms only. Even in Model 3, the ground state which is an anti-ferromagnetic configuration AFM4 (the inter-layer spin configuration is anti-aligned) the magnetic moment of inner surface Fe atoms is 1.15 $ \mu B $, whereas that of the  outer ones is 1.90 $ \mu B $. The itinerant nature of magnetism in Fe is responsible for the larger magnitude of Fe moments in comparison to Ti. 
  
\begin{figure*}
    \includegraphics[height=8cm, width=8.00 cm]{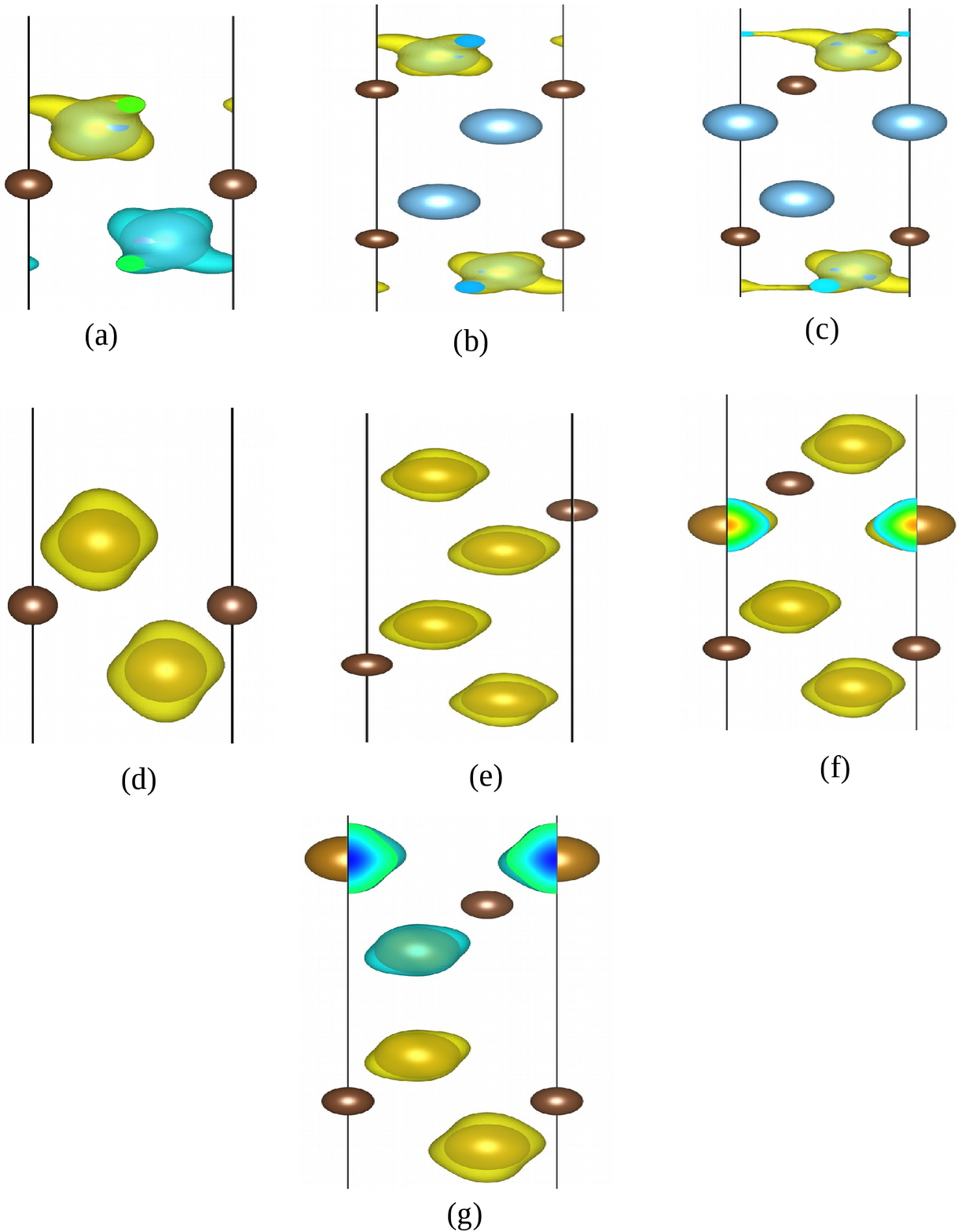}
    \caption{Spin density profile for (a) monolayer Ti\textsubscript{2}C ,  bilayer Ti\textsubscript{2}C (b) Model 1 and (c) Model 2 ,(d) monolayer Fe\textsubscript{2}C ,bilayer Fe\textsubscript{2}C (e) Model 1, (f) Model 2 and (g) Model 3.}
    \label{Fig:3}
\end{figure*}

In order to understand the differences in the magnetism of the two MXenes under consideration, we  plot the spin density profile for both systems in different stacking patterns. The ground state corresponding to each stacking pattern is considered.The results are shown in 
Figure ~\ref{Fig:3}. The results suggest that the spin density in bilayer Ti\textsubscript{2}C is very different from that in the monolayer. Unlike the case of monolayer spin density is completely localised on the outer surface Ti atoms of bilayer Ti\textsubscript{2}C. This explains the reason behind vanishing moments on the inner surface Ti atoms. In contrast,  Fe\textsubscript{2}C there is significant spin densities on both the inner and outer surface Fe atoms in Fe$_{2}$C bi-layers. This is an artefact of the itinerant nature of Fe moments. However, there is noticeable difference in the distributions of the spin density as the stacking pattern changes. 
\subsection{Electronic Structure}
\begin{figure*}
    \includegraphics[height=3cm, width=8.00 cm]{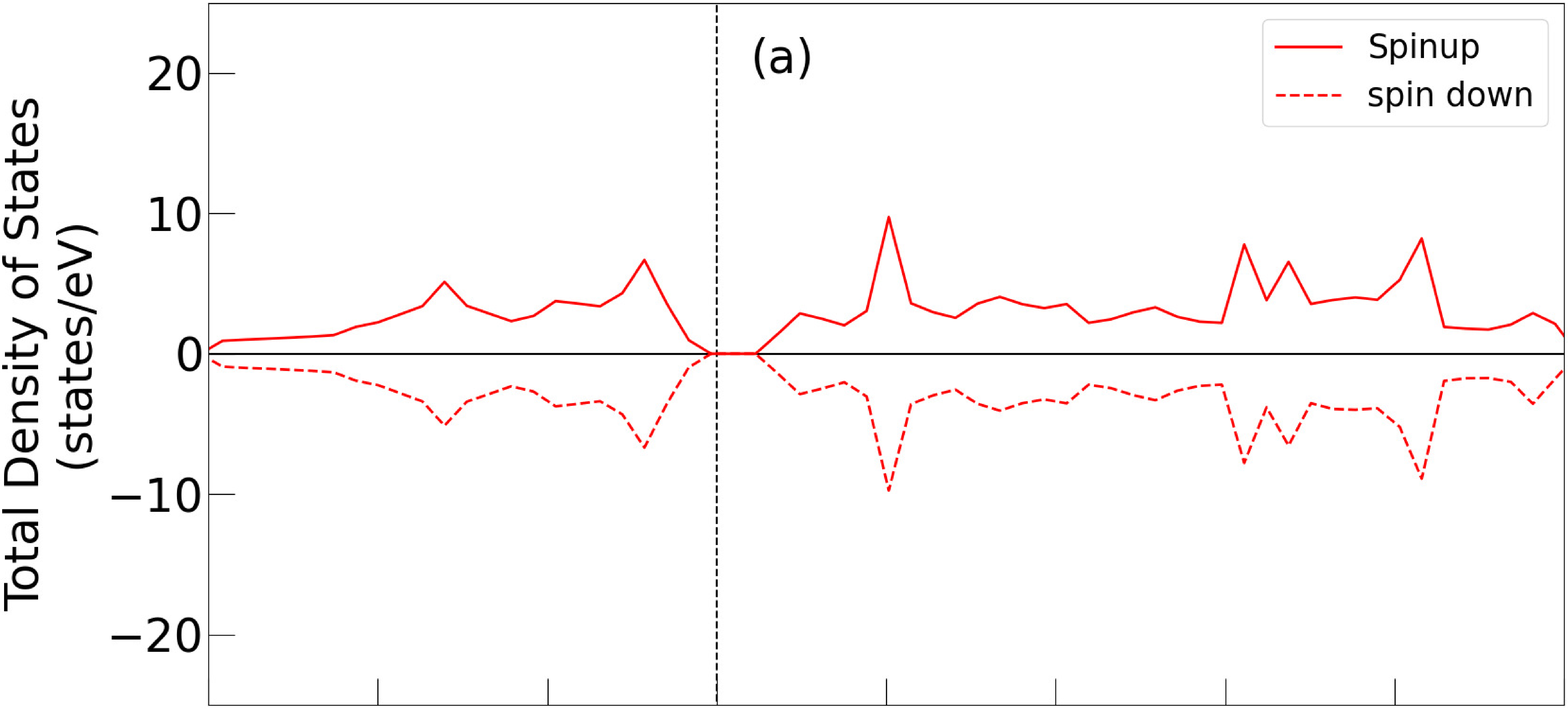}
    \vspace{0.002cm}  
    \includegraphics[height=3cm, width=8.00 cm]{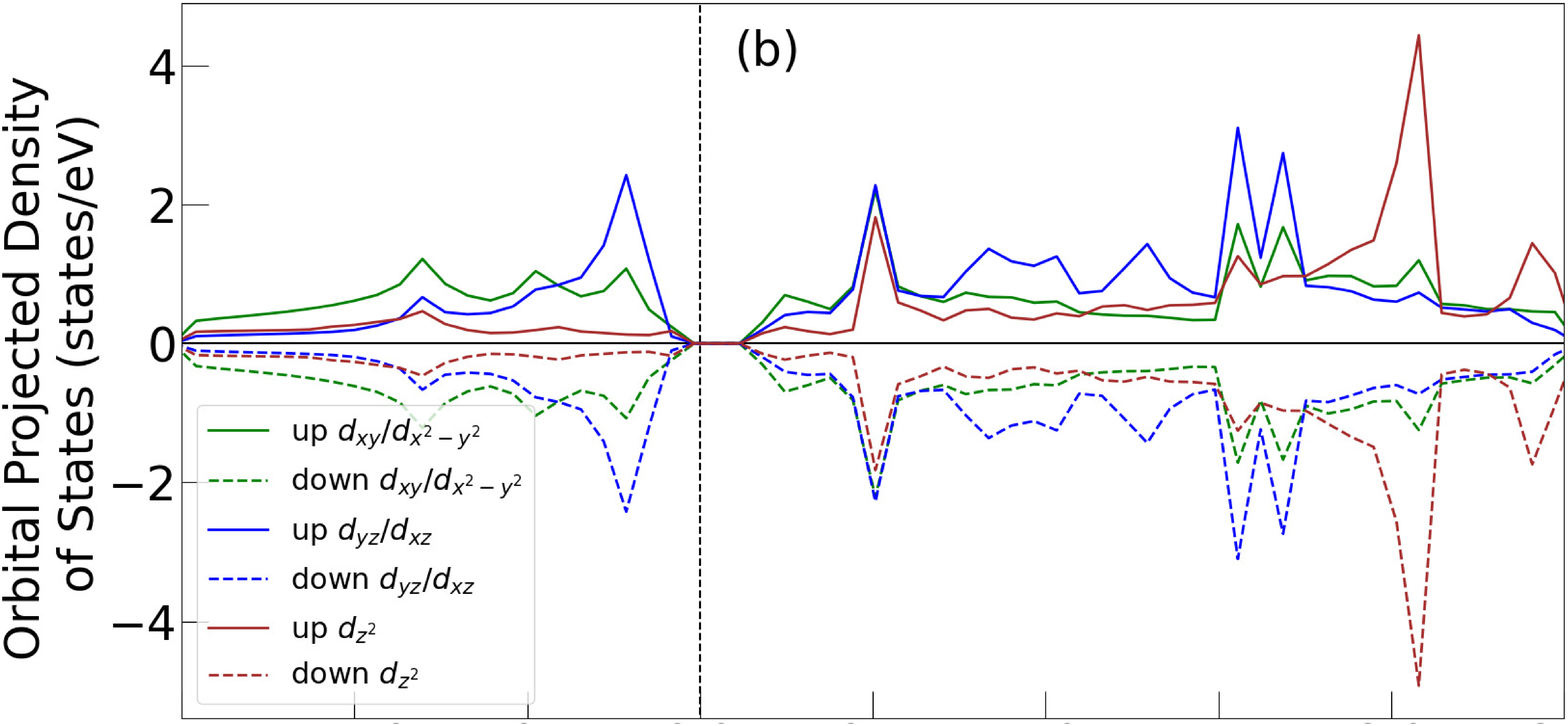} 
    \vspace{0.002cm}
    \includegraphics[height=3cm, width=8.00 cm]{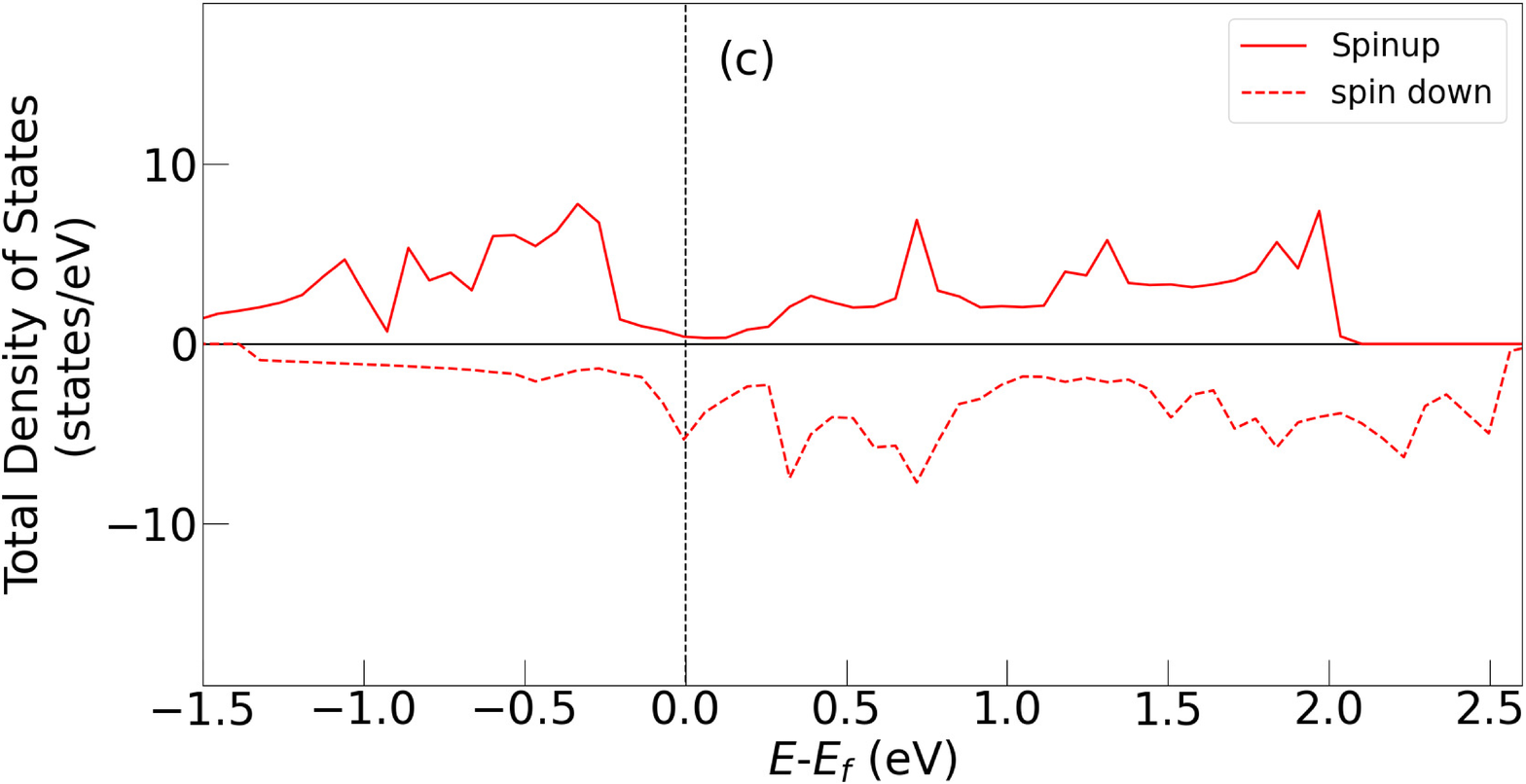}       
    \vspace{0.002cm}
    \includegraphics[height=3cm, width=8.00 cm]{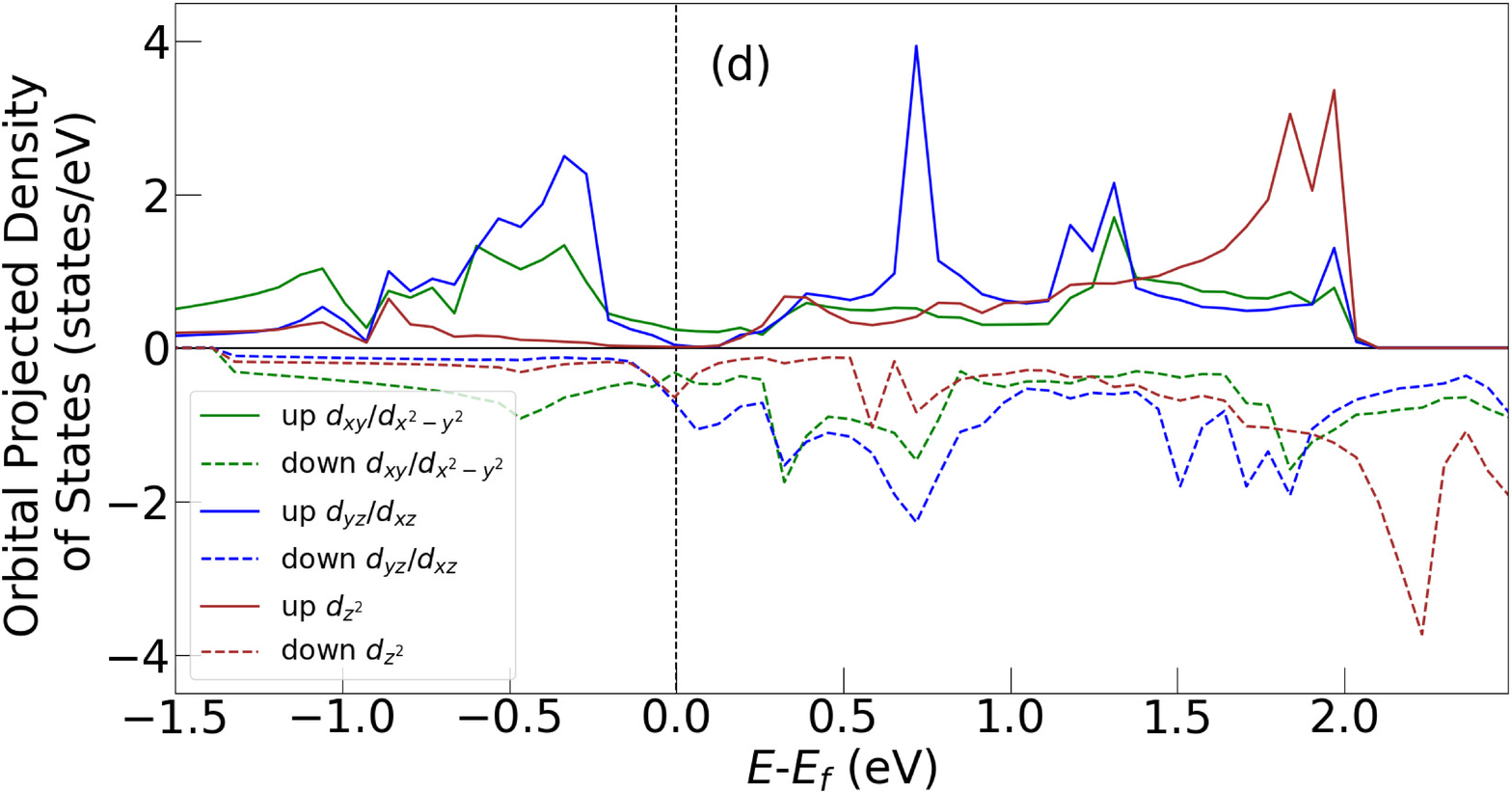}      
     \caption{(a) Total density of states and (b) $d$- orbital projected Ti density of states of monolayer Ti$_{2}$C in ground state A-AFM configuration  are shown. (c) Total density of states  and (d) $d$-orbital projected Ti density of states in FM configuration are shown for comparison.}
    \label{Fig:4}
\end{figure*}
\begin{figure*}
    \includegraphics[height=3cm, width=8.00 cm]{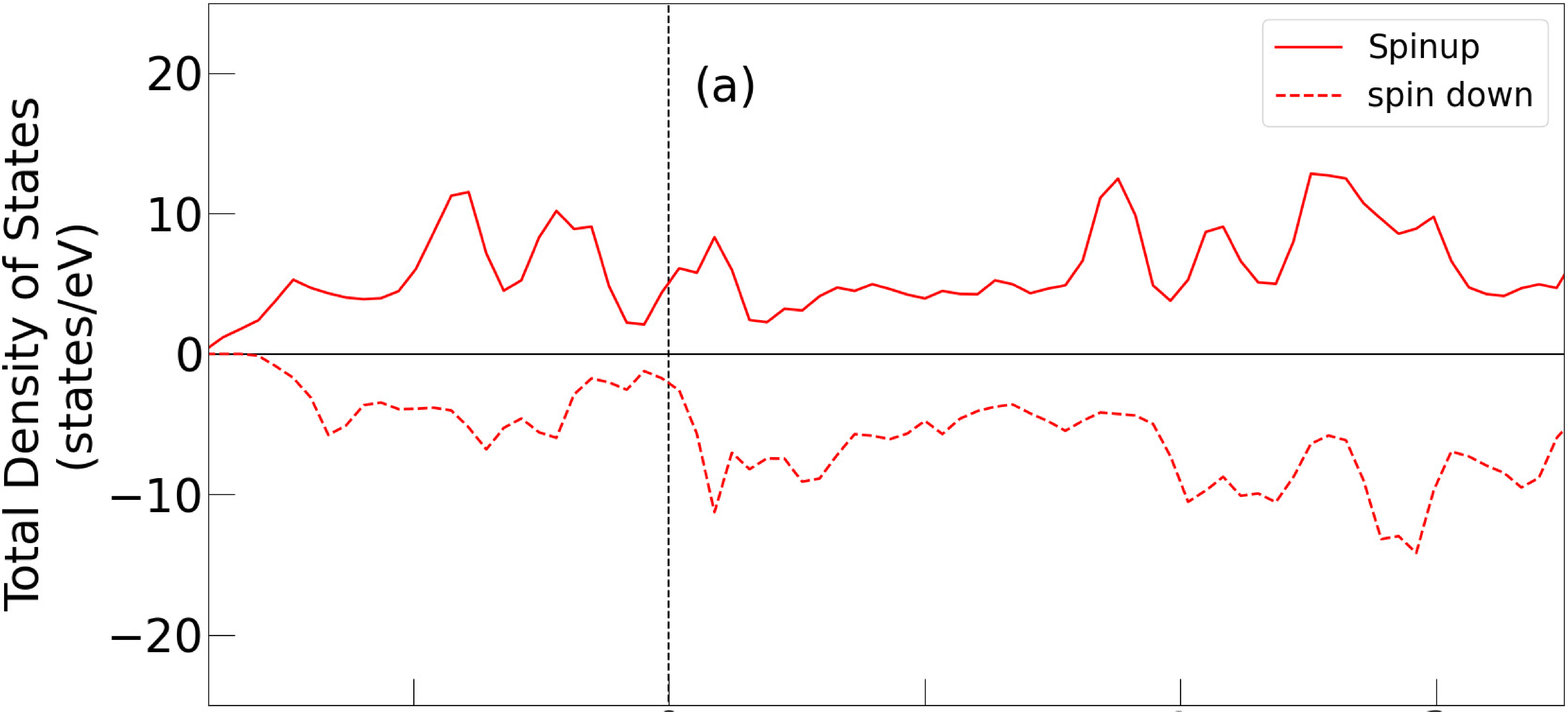}
    \vspace{0.002cm}  
    \includegraphics[height=3cm, width=8.00 cm]{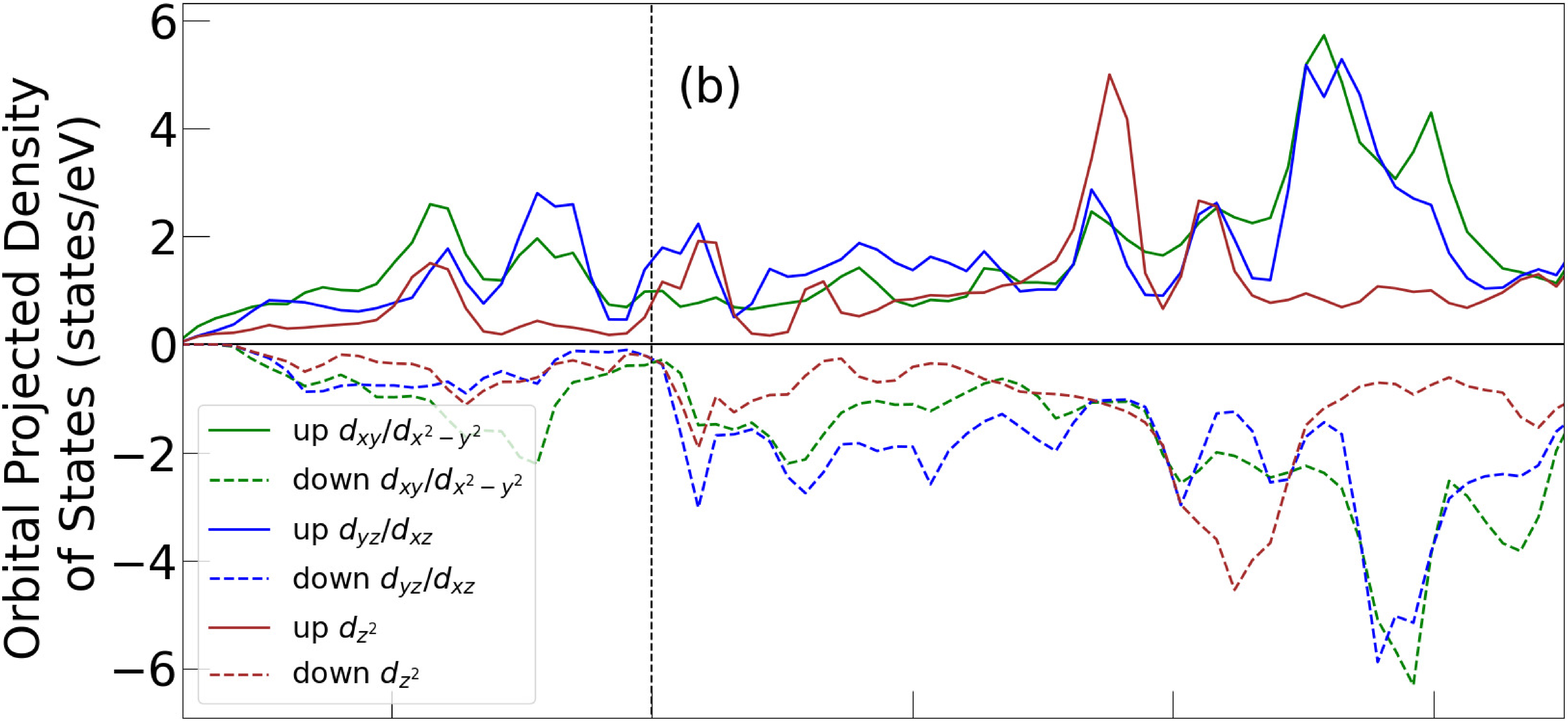} 
    \vspace{0.002cm}
    \includegraphics[height=3cm, width=8.00 cm]{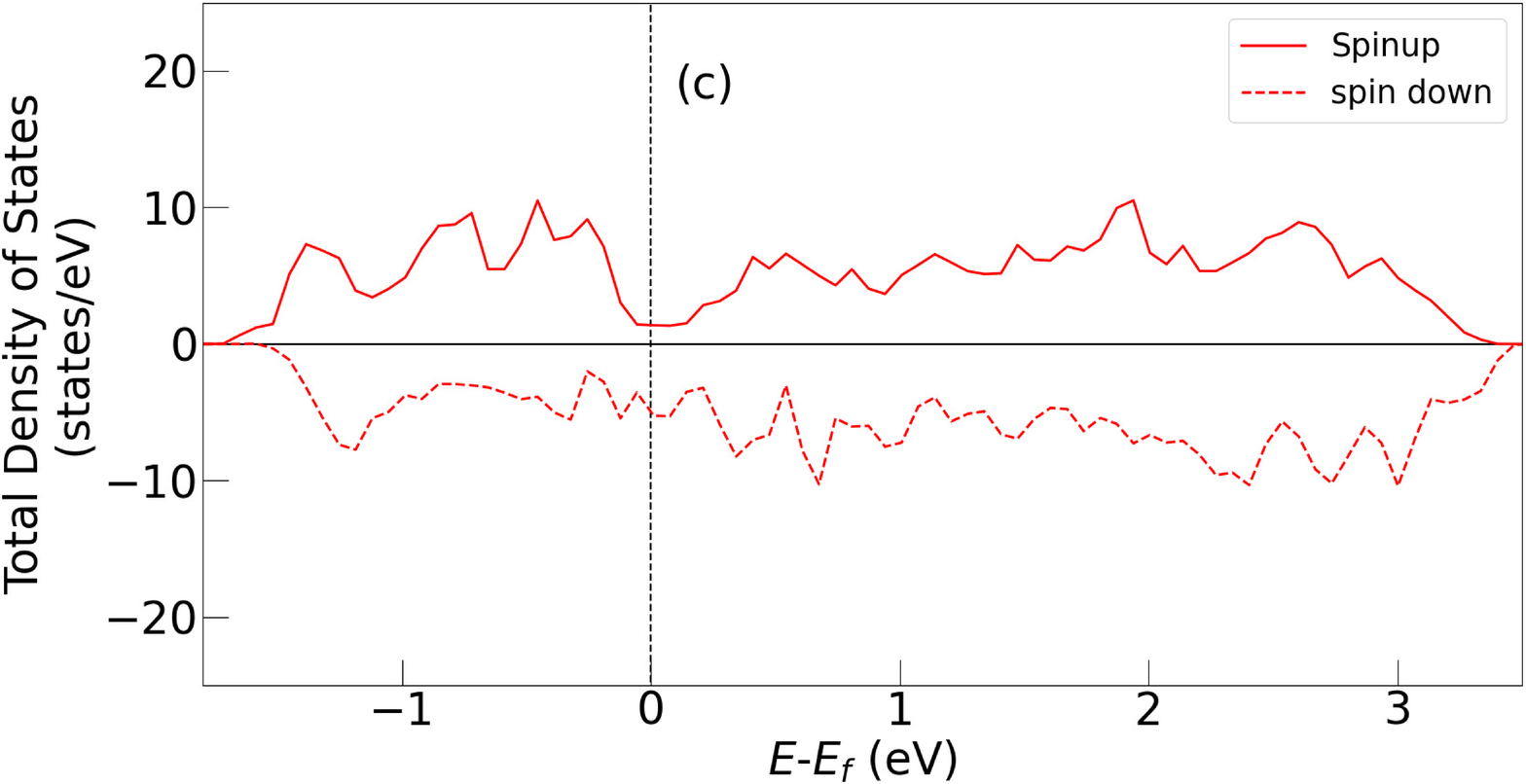}       
    \vspace{0.002cm}
    \includegraphics[height=3cm, width=8.00 cm]{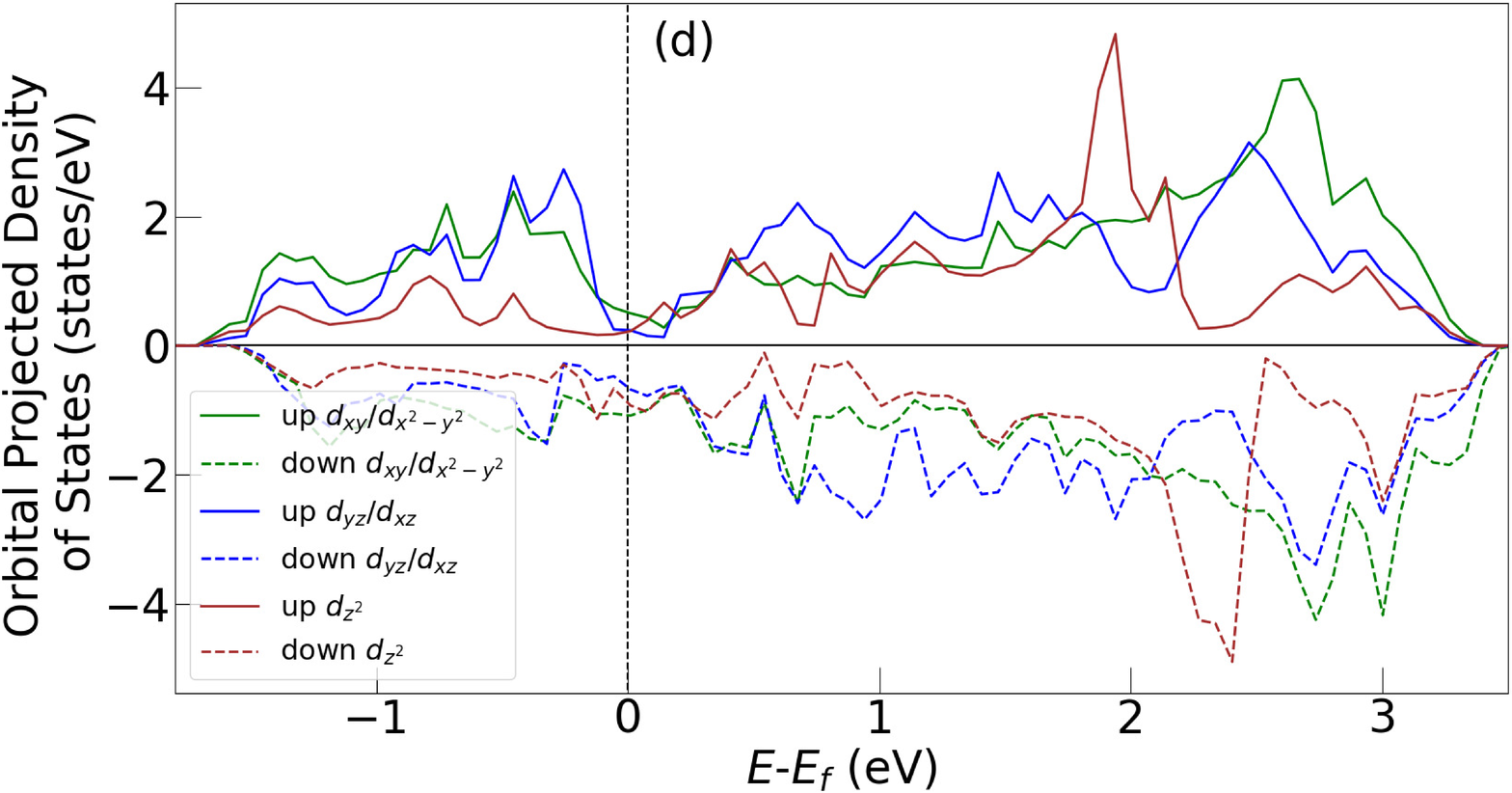}  
    \vspace{0.002cm}
    \includegraphics[height=3cm, width=8.00 cm]{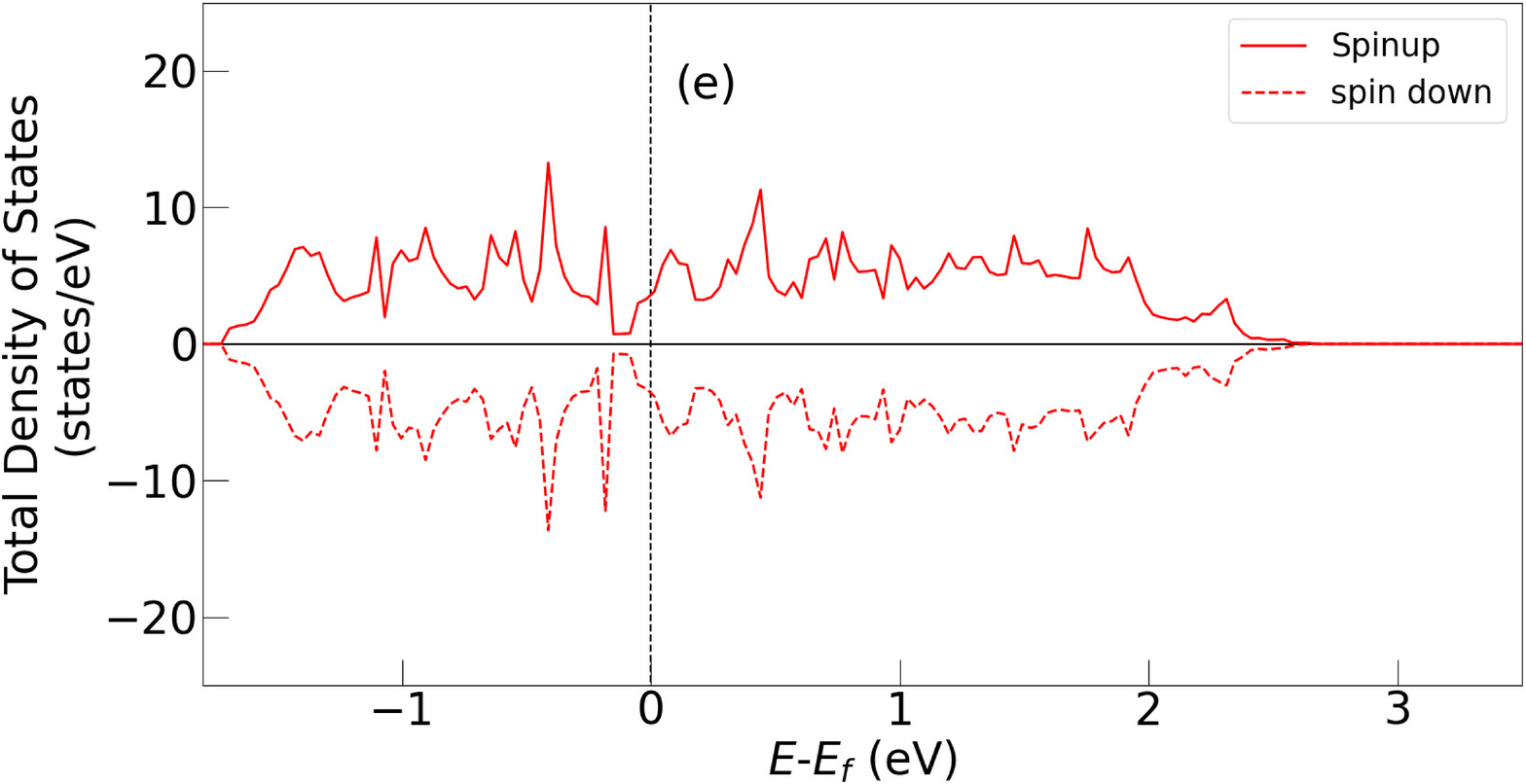}    
    \caption{Total density of states of Ti\textsubscript{2}C bilayer (a) Model 1 and (c) Model 2 in their ground states (FM) are shown.Corresponding $d$-orbital projected Ti density of states in bilayer (b) Model 1 and (d) Model 2 are also shown. Density of states  of bilayer Ti\textsubscript{2}C in AFM1 configuration is shown in (e) for comparison.}
    \label{Fig:5}
\end{figure*}

The total density of states(DOS), and $d$-orbital projected Ti density of states(PDOS) of Ti\textsubscript{2}C monolayer in ground state configuration are shown in Figure ~\ref {Fig:4}(a)-(b). The corresponding bandstructure is shown in Figure S2(a), supplementary information. The results suggest that Ti\textsubscript{2}C monolayer is an indirect band gap semiconductor with a small gap of around 0.22 eV.  In the case of monolayer Ti\textsubscript{2}C, the crystal field with C\textsubscript{3v} symmetry splits the $d$ orbitals of Ti atoms into three states:  d\textsubscript{z\textsuperscript{2}}  and two two-fold degenerate d\textsubscript{x\textsuperscript{2}-y\textsuperscript{2}}/d\textsubscript{xy} and d\textsubscript{xy}/d\textsubscript{yz} state\cite{he2016new}. Figure 4(b) shows that states around the fermi levels are formed by hybridisation among these states of the Ti atom. When bonded with C atoms, each Ti atom donates two 4s electrons to them and becomes Ti$^{2+}$. The remaining two $d$ electrons of each Ti$^{2+}$ ion will half-fill the two-fold degenerate orbitals. This is evident from Figure ~\ref{Fig:4}(b), Figure S5, supplementary information, and Table I of supplementary information. Superexchange interaction between two half-filled $d$-orbitals via non-magnetic Carbon $p$ orbitals, thus, leads to the AFM ground state.  

Figure ~\ref{Fig:5} (a)-(d) show the DOS and PDOS of bilayer Ti\textsubscript{2}C with Model 1 and  2 stacking. Corresponding band structures are shown in Figures S2 (b)-(c), supplementary information. With inclusion of another layer, the gap in the monolayer fills up leading to a semiconductor to metal transformation. It is the states from the outer surface Ti atoms that contribute to the filling of the states around the gap (Figures S6-S7, supplementary information). Information on occupancies in the $d$-orbitals (Table I, supplementary information) suggest that it hardly changes from monolayer to bilayer outer surface Ti atoms. This is irrespective of the stacking pattern. This in turn corroborates the presumption that the magnetism in Ti$_{2}$C is driven by the dangling bonds. The absence of dangling bonds in the Ti atoms located on the inner surfaces coupled with the fact that Ti is weakly magnetic, leads to a near vanishing spin polarisation on them. In order to find out whether the magnetic configuration or the inclusion of one extra layer is responsible for the semiconductor to metal transition, the DOS and PDOS of Ti\textsubscript{2}C monolayer in FM configuration is shown in Figure ~\ref{Fig:4}(c)-(d). We find that in this magnetic configuration, the monolayer Ti$_{2}$C is like a semi-metal with small densities of states around the gap in AFM configuration. It is, therefore, tempting to conclude that the changes in the magnetic configuration drives the changes in the electronic structure. To cross-confirm, we look at the DOS of bilayer Ti\textsubscript{2}C with Model 2 stacking in AFM1 magnetic configuration (Figure ~\ref{Fig:5}(e)). The AFM1 configuration is chosen as it is the closest to A-AFM configuration in monolayer structure. We find that even with this AFM spin configuration, the bilayer structure is metallic in nature. This implies that if the  magnetic configuration was solely responsible for the changes in the electronic structure, we would have gotten AFM semiconductor in bilayer Model 2 stacking as well. We, therefore, conclude that the presence of one extra layer must be responsible for the metallicity in Ti\textsubscript{2}C bilayer. From the PDOS plots (Figure ~\ref{Fig:5}(b),(d) and Figure S6-7, supplementary information) it is observed that the states near Fermi levels are mainly contributed by the d\textsubscript{xy}/d\textsubscript{x\textsuperscript{2}-y\textsuperscript{2}} orbitals of Ti atoms of both layers and $s$ states of the outer surface Ti atoms.
\begin{figure*}
\includegraphics[height=3cm, width=8.00 cm]{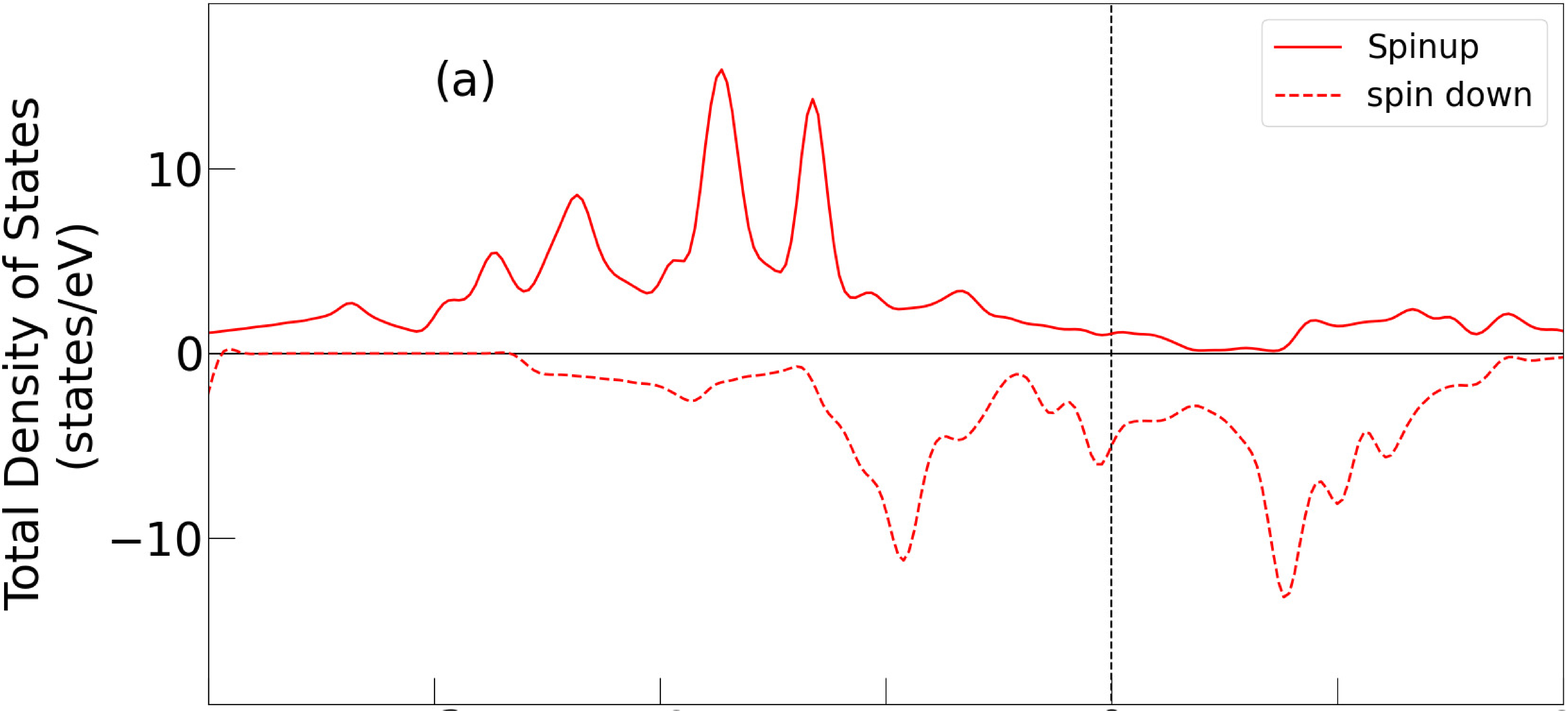}
    \hspace{0.2cm}
    \includegraphics[height=3cm, width=8.00 cm]{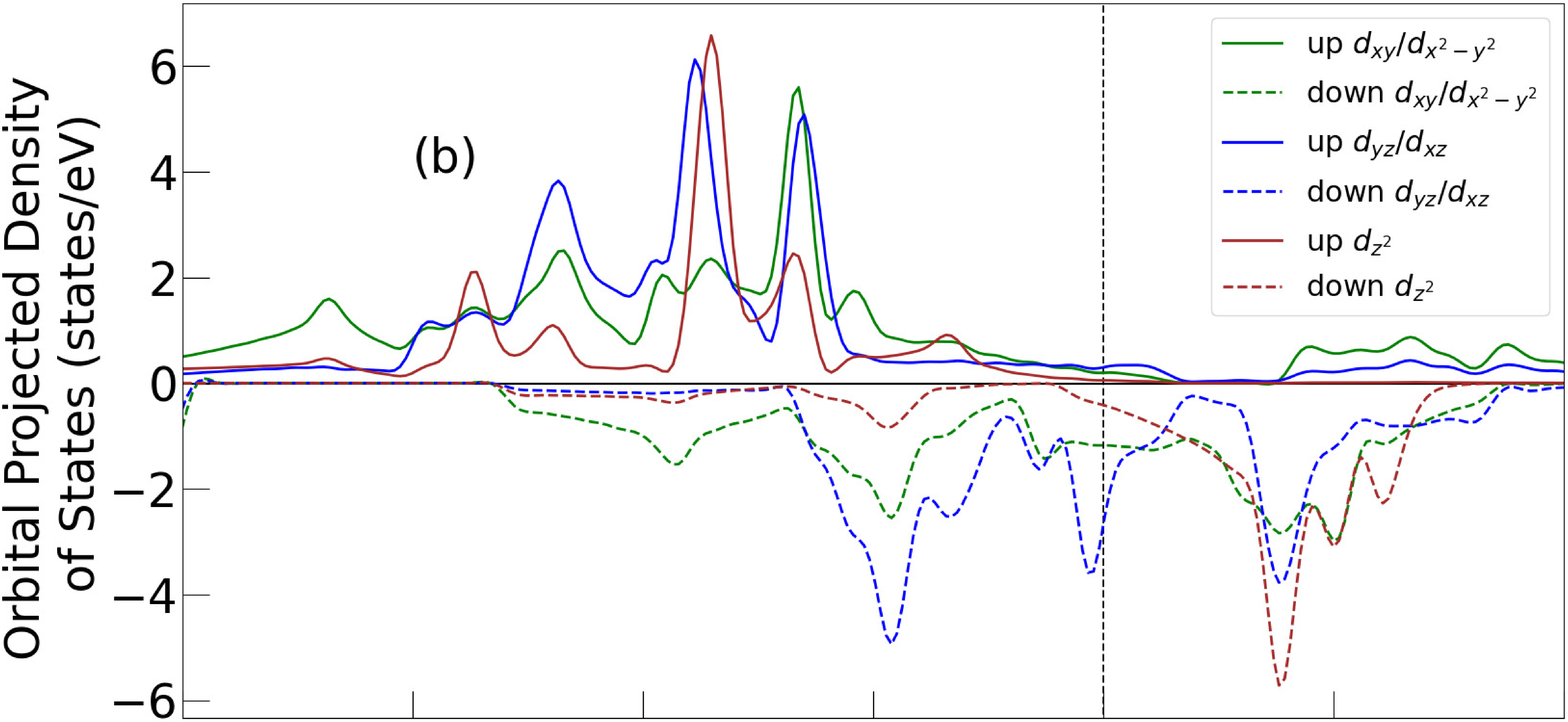}  
    \vspace{0.002cm}
    \includegraphics[height=3cm, width=8.00 cm]{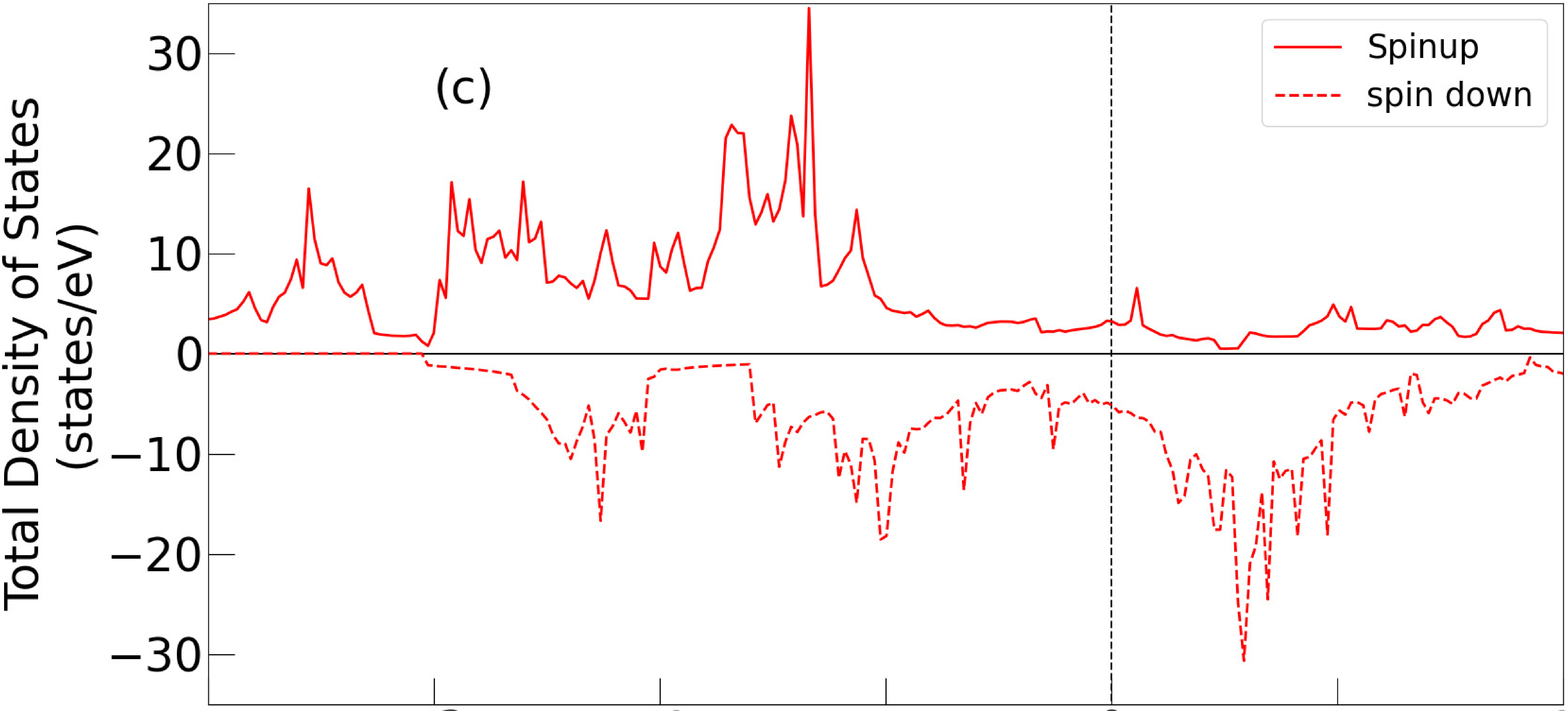}
    \hspace{0.35cm}  
    \includegraphics[height=3cm, width=8.00 cm]{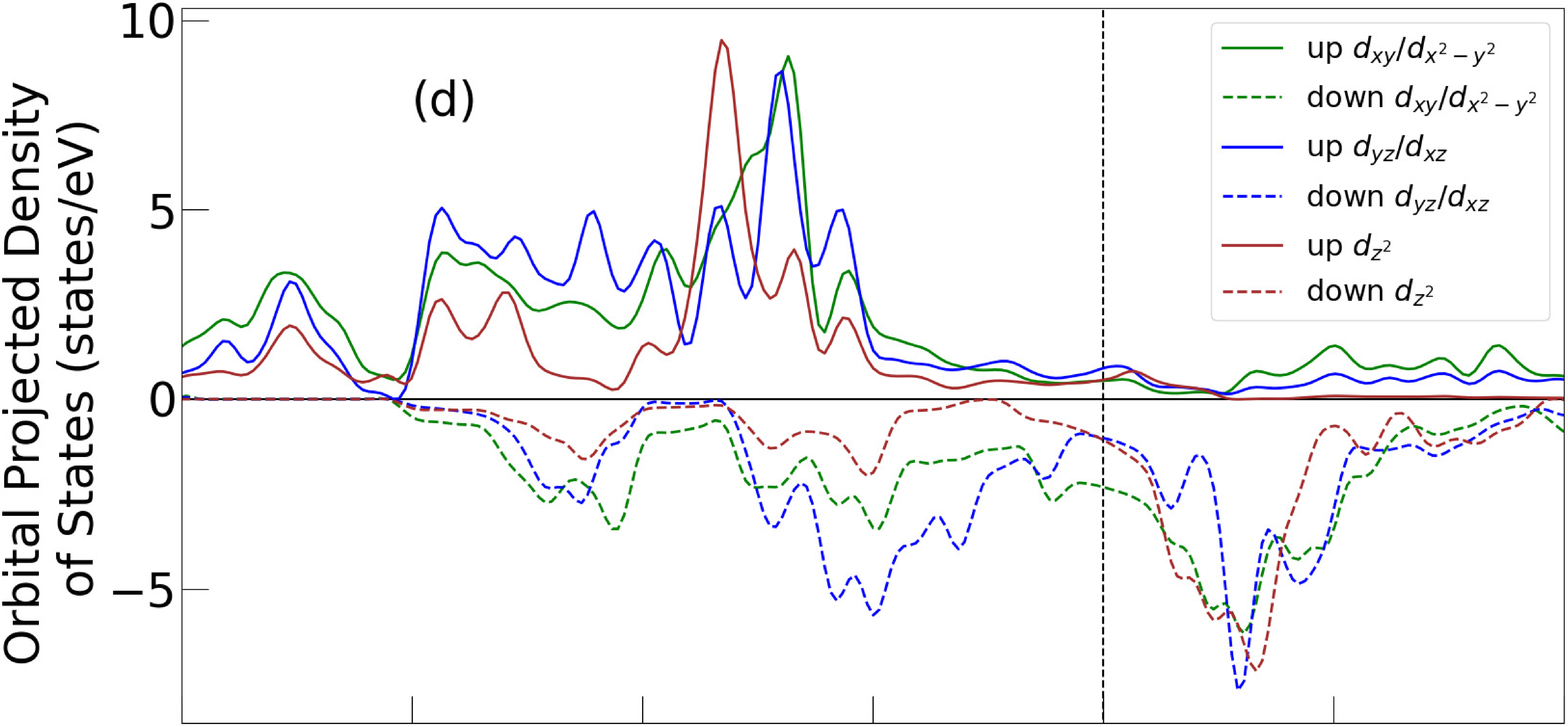} 
    \vspace{0.002cm}
    \includegraphics[height=3cm, width=8.00 cm]{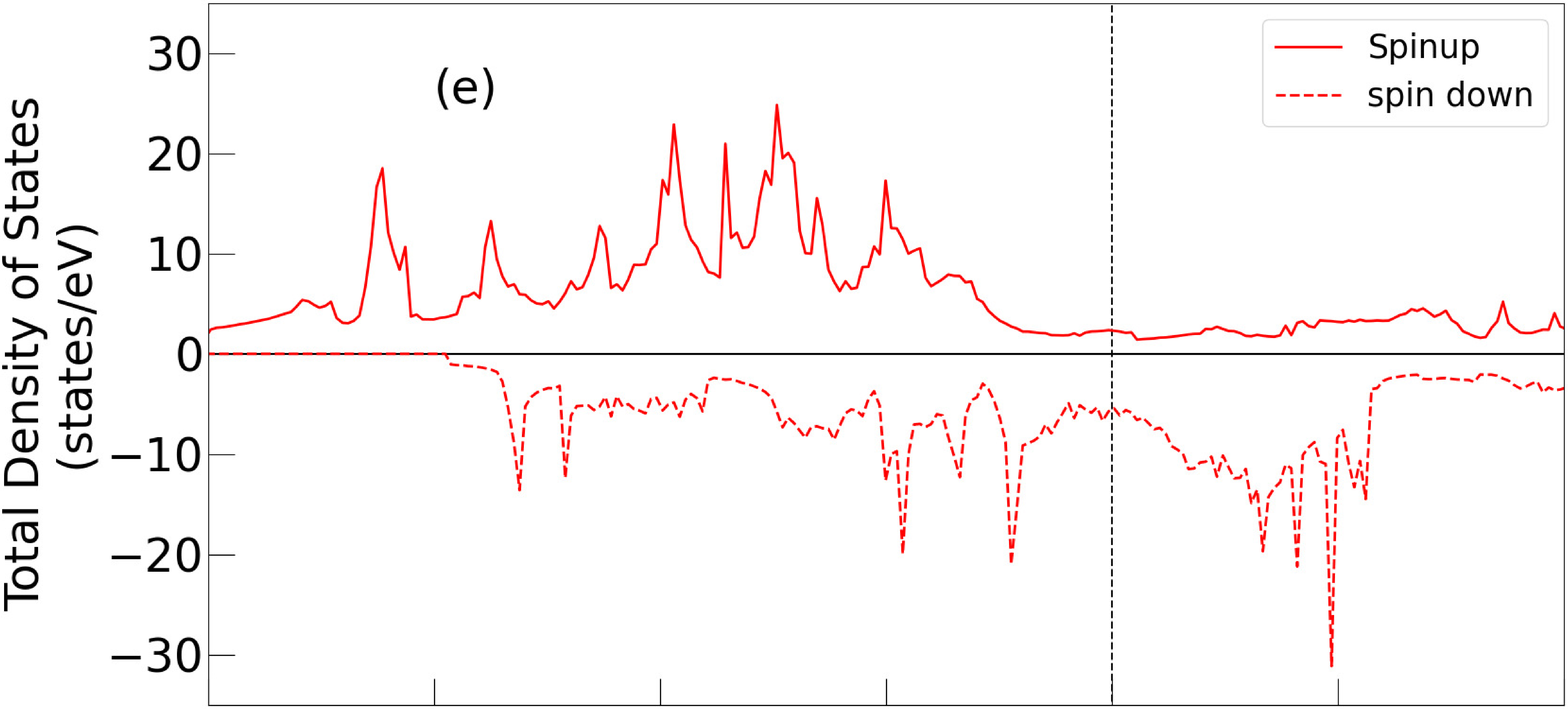}       
    \hspace{0.355cm}
    \includegraphics[height=3cm, width=8.00 cm]{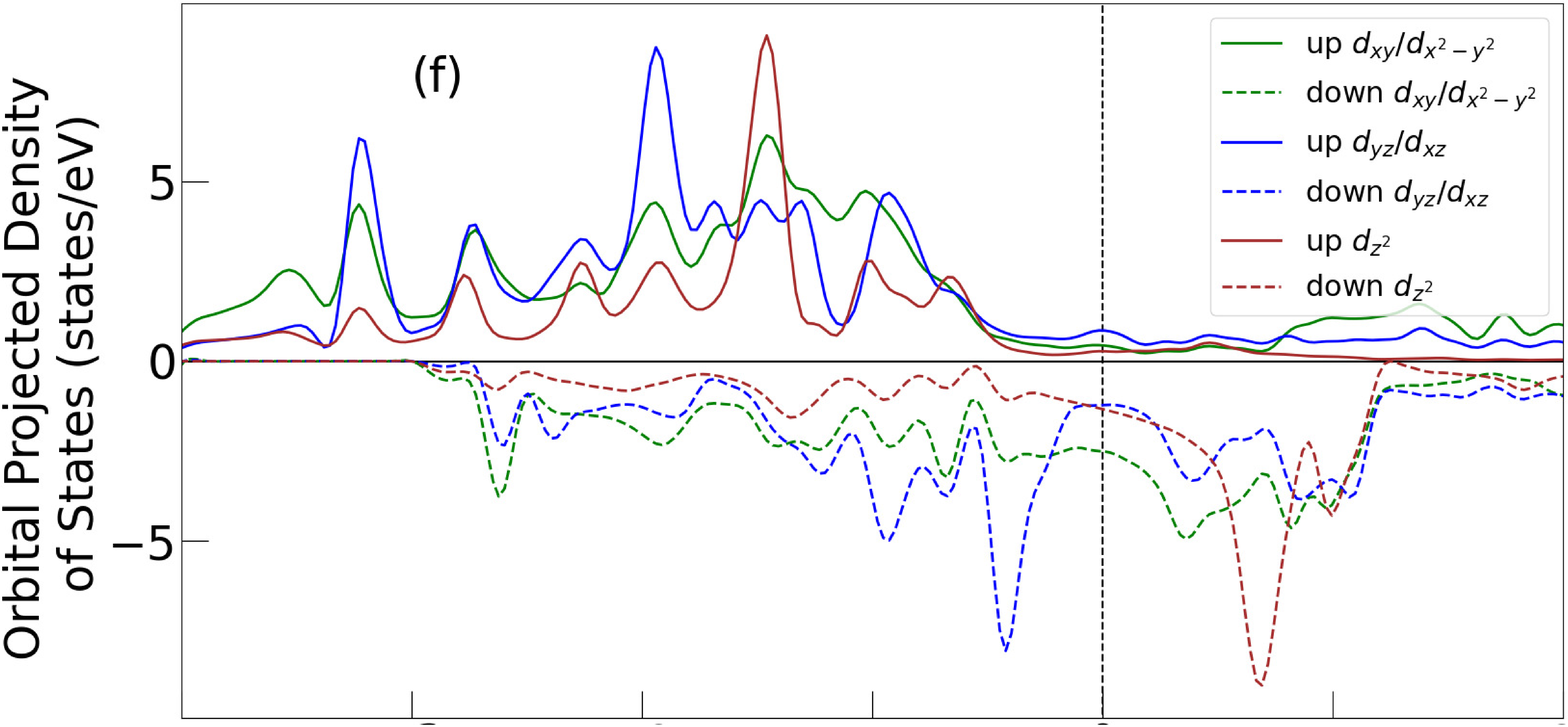}  
    \vspace{0.002cm}
    \includegraphics[height=3cm, width=8.00 cm]{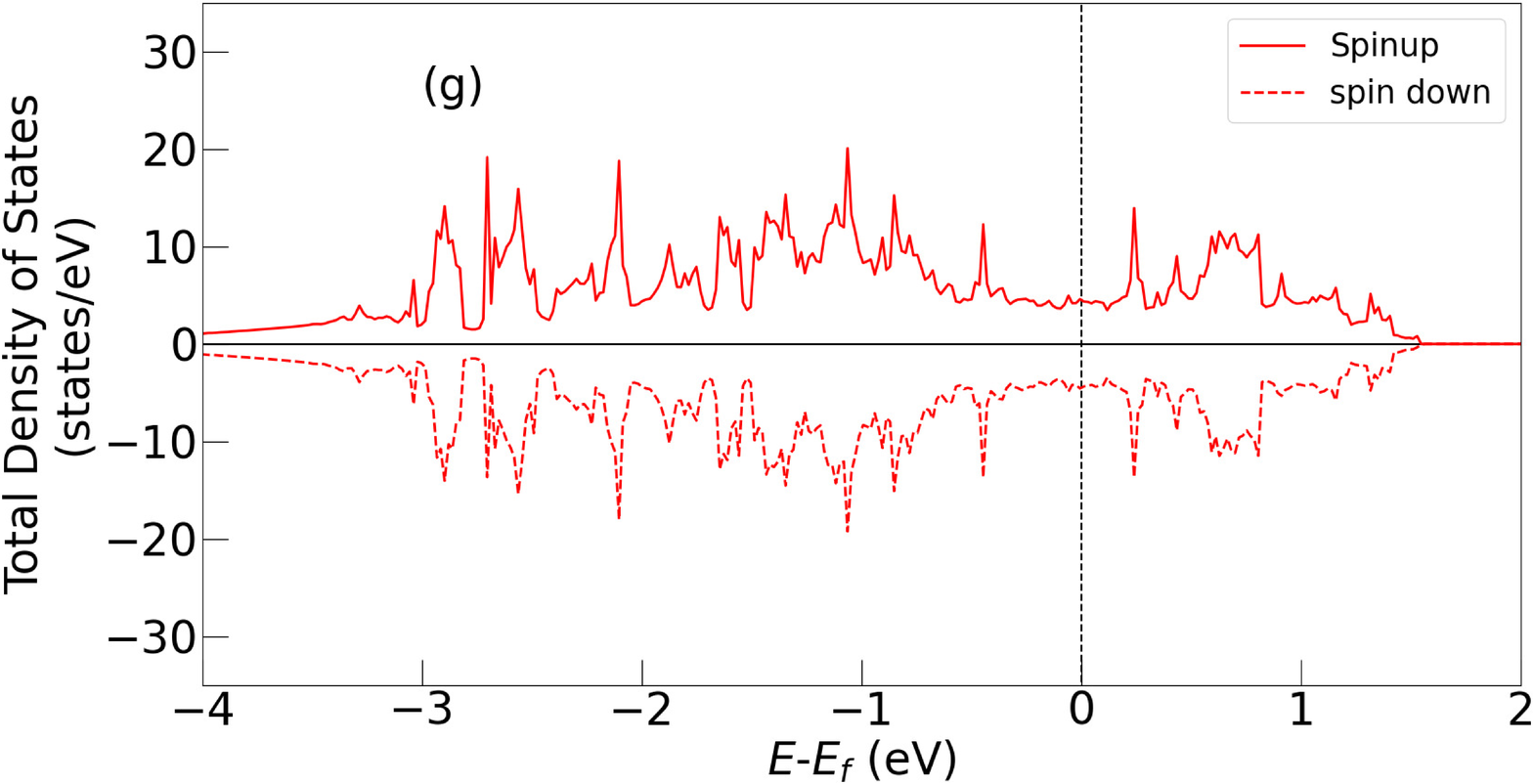}
    \hspace{0.35cm}
    \includegraphics[height=3cm, width=8.00 cm]{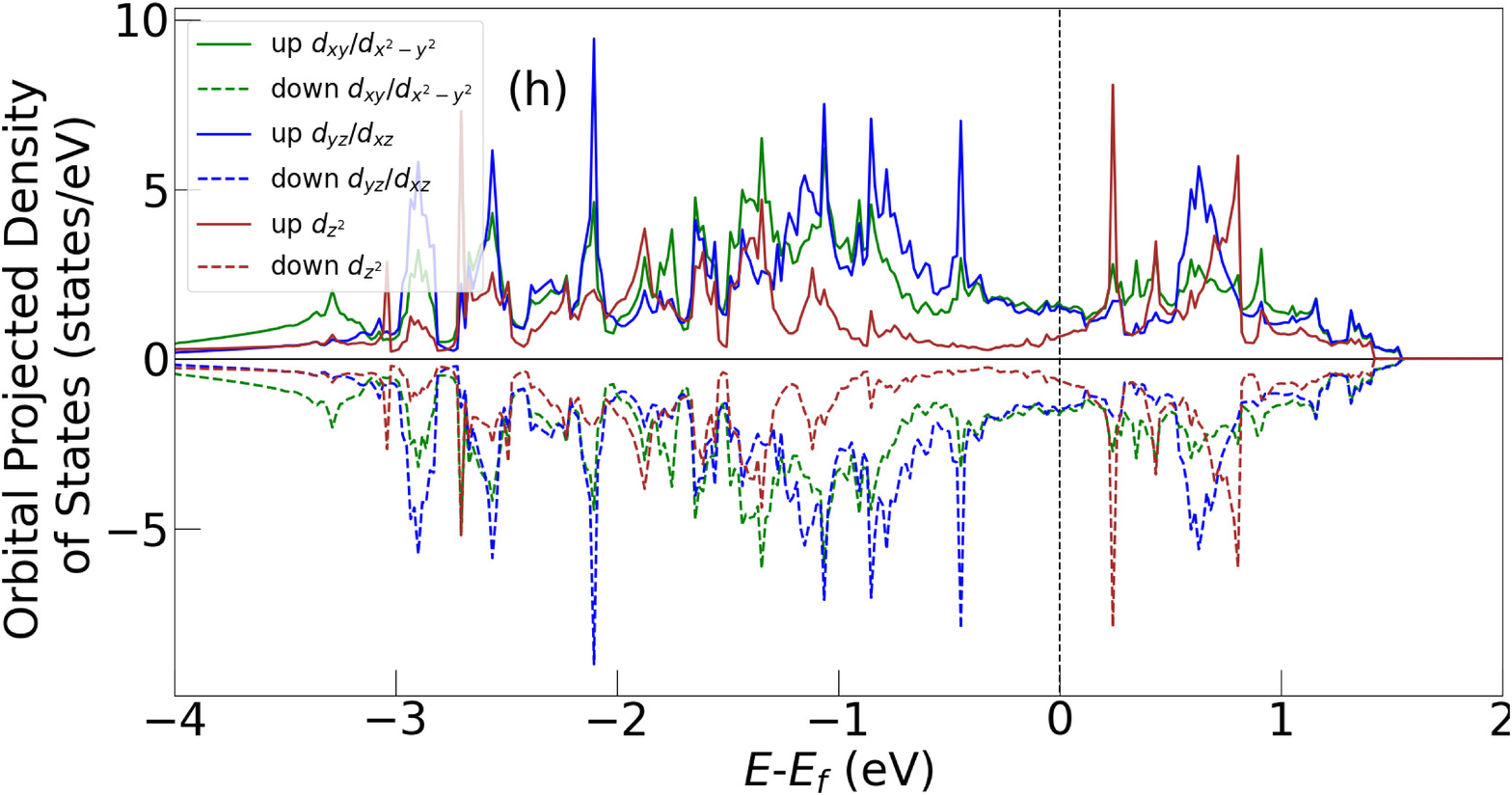}     
    \caption{Total density of states of Fe\textsubscript{2}C (a) monolayer and bilayer (c)Model 1, (e) Model 2 and (g) Model 3 are shown. $d$-orbital projected Ti density of states for monolayer and bilayer Model 1,Model 2 and Model 3 are shown in (b),(d),(f) and (h),respectively. }
    \label{Fig:6}
\end{figure*}
In Figure ~\ref{Fig:6}(a)-(b) and Figure S2 (d), supplementary information, we show the DOS, PDOS, and band structures of monolayer Fe\textsubscript{2}C. DOS of bilayer Model 1, Model 2 and Model 3 are shown in Figure ~\ref{Fig:6} (c), (e) and (f), respectively. The PDOS in these three models are displayed in Figure ~\ref{Fig:6} (d),(f) and (h). The corresponding band structures are shown in Figure S2 (e)-(f), supplementary information. Irrespective of the stacking, the system is metallic with states near Fermi levels are mainly due to hybridization among the d\textsubscript{xy}/d\textsubscript{x\textsuperscript{2}-y\textsuperscript{2}} and d\textsubscript{yz}/d\textsubscript{zx} orbitals. Unlike Ti$_{2}$C, in case of bilayer Model 1 and Model 2 stacking in Fe$_{2}$C, both outer surface and inner surface Fe atoms are significantly spin polarised as is obvious from Figures S9-S10, supplementary information. There are, however, noticeable differences in the electronic re-distribution among the spin bands between the two stackings. While there is not much change in the occupations in the spin bands of outer surface Fe atoms, in Model 2, the spin down band of inner surface Fe atom is more populated than that of Model 1 at the expense of its spin up band. This leads to the differences in the atomic magnetic moments in these two models. The electron numbers shown in Table I, supplementary information, corroborates this. The AFM ground state in case of Model 3 makes the electronic structure very different (Figure ~\ref{Fig:6}(g)-(h), Figure S2 (g) and Figure S11, supplementary information). This must be due to differences in the magnetic interactions as a consequence of differences in the stackings. In the next sub-section we will address this. 
\subsection{Magnetic Exchange  parameters and  transition temperatures}
\begin{table*}
\caption{\label{tab:Table4}Magnetic Exchange parameters, Curie/Neels Temperature and MAE of monolayer and bilayer MXenes}
\begin{tabular}{ m{0.2\textwidth}m{0.1\textwidth}m{0.1\textwidth}m{0.14\textwidth}m{0.15\textwidth}m{0.1\textwidth}}
 \hline
 Material & J\textsubscript{1}(meV) & J\textsubscript{2}(meV) & J\textsubscript{3}(meV) & Curie/Neel Temperature(K) & MAE ($ \mu$eV/atom)  \\
 \hline
 Ti\textsubscript{2}C Monolayer & -5.14 & 10.33 & -1.01 & 125   & 5.68\\ 
 
 Ti\textsubscript{2}C Bilayer Model 1 & 1.33 & 1.16 & 4.69  & 185  & 4.74\\ 
 
 Ti\textsubscript{2}C Bilayer Model 2 & -2.25 & 1.27 & 4.08 & 175  & 2.45\\ 
 \hline
 
 Fe\textsubscript{2}C Monolayer & 5.27 & 4.65 & 0.87 & 450  & -7.02 \\
 
 Fe\textsubscript{2}C Bilayer Model 1 & 9.28 & 7.11 & 5.43 & 625  & 1.82\\
 
 Fe\textsubscript{2}C Bilayer Model 2 & 4.63  & 7.94 & 1.78 & 410  & -14.57\\
 
 Fe\textsubscript{2}C Bilayer Model 3 & -3.92 & 0.95 & 4.85 & 380  & 41.5\\
 \hline
\end{tabular}
\end{table*}

In order to calculate the magnetic exchange interactions, we make use of the 2D Ising model with first three M-M exchange interactions $J_{1}, J_{2}$ and $J_{3}$ \cite{luo2020first}:
\begin{equation}
H = -\sum_{ij}J_{1}m_{i}m_{j} -\sum_{kl}J_{2}m_{k}m_{l} - \sum_{mn}J_{3}m_{m}m_{n}
\label{Eqn:1}
\end{equation}
 For the monolayers, mapping of energies of  the four magnetic configurations considered, to the 2D Ising Hamiltonian yields the following set of equations:
\begin{equation}
E_{FM} = E_{0}-6J_{1}m^{2}-12J_{2}m^{2}-6J_{3}m^{2}
\label{Eqn:2}
\end{equation}
\begin{equation}
E_{A-AFM} = E_{0}+6J_{1}m^{2}-12J_{2}m^{2} + 6J_{3}m^{2}
\label{Eqn:3}
\end{equation}
\begin{equation}
 E_{C-AFM} = E_{0}-2J_{1}m^{2}+4J_{2}m^{2} + 6J_{3}m^{2}
\label{Eqn:4}
\end{equation}
\begin{equation}
E_{G-AFM} = E_{0}+2J_{1}m^{2}+4J_{2}m^{2} - 6J_{3}m^{2}
\label{Eqn:5}
\end{equation}
\begin{figure}
    \includegraphics[height=4.4 cm, width=7.00 cm]{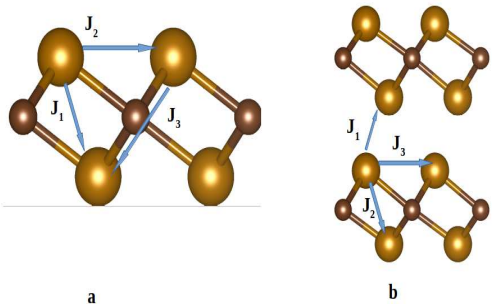}
 \caption{First three magnetic exchange interaction parameters J\textsubscript{1},J\textsubscript{2} and J\textsubscript{3} in monolayer Fe\textsubscript{2}C and bilayer Fe\textsubscript{2}C Model 1 are shown.Golden balls are the Fe atoms and brown ones are the C atoms}
    \label{Fig:7} 
\end{figure}
The $J_{i}$s are obtained by solving the set of equations.
Extension of  the procedure to include the eight magnetic configurations considered provides us with the exchange interactions for bilayers. These interactions are then used to calculate the magnetic transition temperatures by classical Monte Carlo simulations. The magnetic exchange interactions and the transition temperatures calculated this way for Ti$_{2}$C and Fe$_{2}$C are shown in Table ~\ref{tab:Table4}.

The results for Ti$_{2}$C monolayer shows comparable strengths of first and second neighbour exchange interactions. While J$_{1}$ signifies AFM interaction, J$_{2}$ implies FM interaction. The AFM first neighbour interaction is superexchange one via the C atoms coordinated between the two Ti atoms (Figure S3 (a)) while the second neighbour interaction is direct exchange. As has been discussed in Section IIIB, the superexchange between the half filled Ti orbitals drives the AFM ground state of this system. The calculated magnetic exchange interactions justify this explanation. In case of Fe$_{2}$C monolayer, all three magnetic exchange parameters signify a FM ground state. In this case, the FM state is driven by the first neighbour superexchange between more than half-filled Fe $d$ orbitals. 

The magnetic interactions for bilayers are quite intriguing. For the bilayers considered here, $J_{1}, J_{3}$ are intra-layer while $J_{2}$ is inter-layer in case of Ti$_{2}$C. Things are different for Fe$_{2}$C; in bilayer Models 1 and 3, $J_{1}$ is inter-layer while the other two are intra-layer. The interactions are schematically shown in Figure ~\ref{Fig:7} and Figure S3, supplementary information. In case of Ti$_{2}$C, a change in stacking inflicts very little change in the exchange interactions. The qualitative change happens for $J_{1}$ when it becomes AFM in Model 2 from FM in Model 1. This is due to the competing direct and superexchange that change with orientations of the Ti atoms on the inner and outer surfaces of the bilayer. The presence of an extra layer as compared to monolayer renormalises the two competing exchanges. Clearly the main effect is to  reduce the strength of the C mediated superexchange which is prominent in both models of stacking pattern. Thus in Ti$_{2}$C bilayers FM due to direct exchange dominates. This becomes clearer by looking at the strengths of J$_{3}$. 

Strong FM interactions in either of Model 1 and Model 2 , Fe$_{2}$C bilayer is consistent with the FM ground state obtained by total energy calculations. However, to understand how the stacking affects various interactions, $J_{1}(J_{2})$ of Model 1 is to be compared with $J_{2}(J_{1})$ of Model 2. We find that the strength of the interactions decrease when the stacking changes to Model 2 from from Model 1. The same qualitative feature was found in Ti$_{2}$C. Such trend across systems originates from the effect on the overlap of the spin wave functions due to changes in the orientations of the transition metal atoms due to changes in the stacking. A comparison between the exchange parameters of Model 1 and Model 3 shows overwhelming reduction in the FM interaction for first and second neighbour interactions. This is an artefact of the changes in the competing superexchange and direct exchanges with the changes in the stacking pattern. The AFM $J_{1}$ in Model 3 is consistent with the AFM ground state obtained from total energy calculations. It is noteworthy that $J_{3}$ does not suffer such drastic change across models, presumably due to the dominant direct exchange nature of the interaction.  

The magnetic transition temperatures for both systems follow the trends in the variations in the magnetic exchange interactions. Monolayer Fe$_{2}$C has larger transition temperature in comparison with Ti$_{2}$C monolayer due to stronger FM exchange coupling in case of the former. Since the interactions are predominantly FM in Fe$_{2}$C bilayer, the transition temperatures in general are higher in this system with the highest one of 610 K obtained in Model 1 stacking. The strongest FM exchange interactions in this system is responsible for this. 
 \subsection{Magnetic Anisotropy energy (MAE)}
 From the point of view of applications, one of the important properties of magnetic materials is magnetic anisotropy. Systems with considerable magnetic anisotropy energy (MAE) are usually preferred for various storage-related applications. This MAE makes the existence of magnetic ordering in these classes of materials possible\cite{stohr2006magnetism}. MAE is evaluated the following way:
 \begin{equation}
E_{MAE} = E_{\parallel}-E_{\perp}
\label{Eqn:5}
\end{equation}
$ E_{\parallel} $ and $ E_{\perp} $ are  DFT total energies obtained  by aligning the spins in-plane and out of plane with respect to the surface. Spin-orbit coupling (SOC) is taken to account to calculate these energies. The total energies for all structures are obtained along  the high symmetric in-plane (along 100 and 010 axis) and out-of-plane directions(001). The direction having the lowest energy is chosen as the magnetic easy direction for that particular structure. For example, in Ti\textsubscript{2}C  monolayer (001) direction has the lowest energy; hence this is the magnetic easy axis for monolayer Ti\textsubscript{2}C. The calculated MAE are presented in Table ~\ref{tab:Table4}. For both monolayer and bilayer Ti$_{2}$C, (001) turns out to the magnetisation easy axis. The positive values of MAE too suggests the presence of an easy axis. We find that there is no drastic change in MAE values upon changes in stacking for bilayer Ti$_{2}$C. The change can be attributed to the changes in DOS near the fermi level as stacking changes. 

For Fe\textsubscript{2}C monolayer, the in-plane directions (100) and (010) have the same  energy. The MAE turns out to be negative implying  presence of an easy plane of magnetization. This suggests the magnetism in this system describable by  XY model, and that there is a possibility of a non-collinear ground state.For Fe\textsubscript{2}C bilayer, Model 1 and Model 3, the MAE values are positive indicating the presence of an easy axis for both cases; it turns out to be the (001) axis. The drastic increase in MAE in Model 3 can be attributed to an increase in states in DOS at the fermi level in the majority spin channel. For  Model 2 stacking, MAE value is negative implying the presence of an easy plane instead of an easy axis and thus possibility of a non-collinear ground state.

\begin{figure}
 \center
   \begin{subfigure}{0.50\textwidth}
     \centering
      \includegraphics[height=6cm, width=8.20 cm]{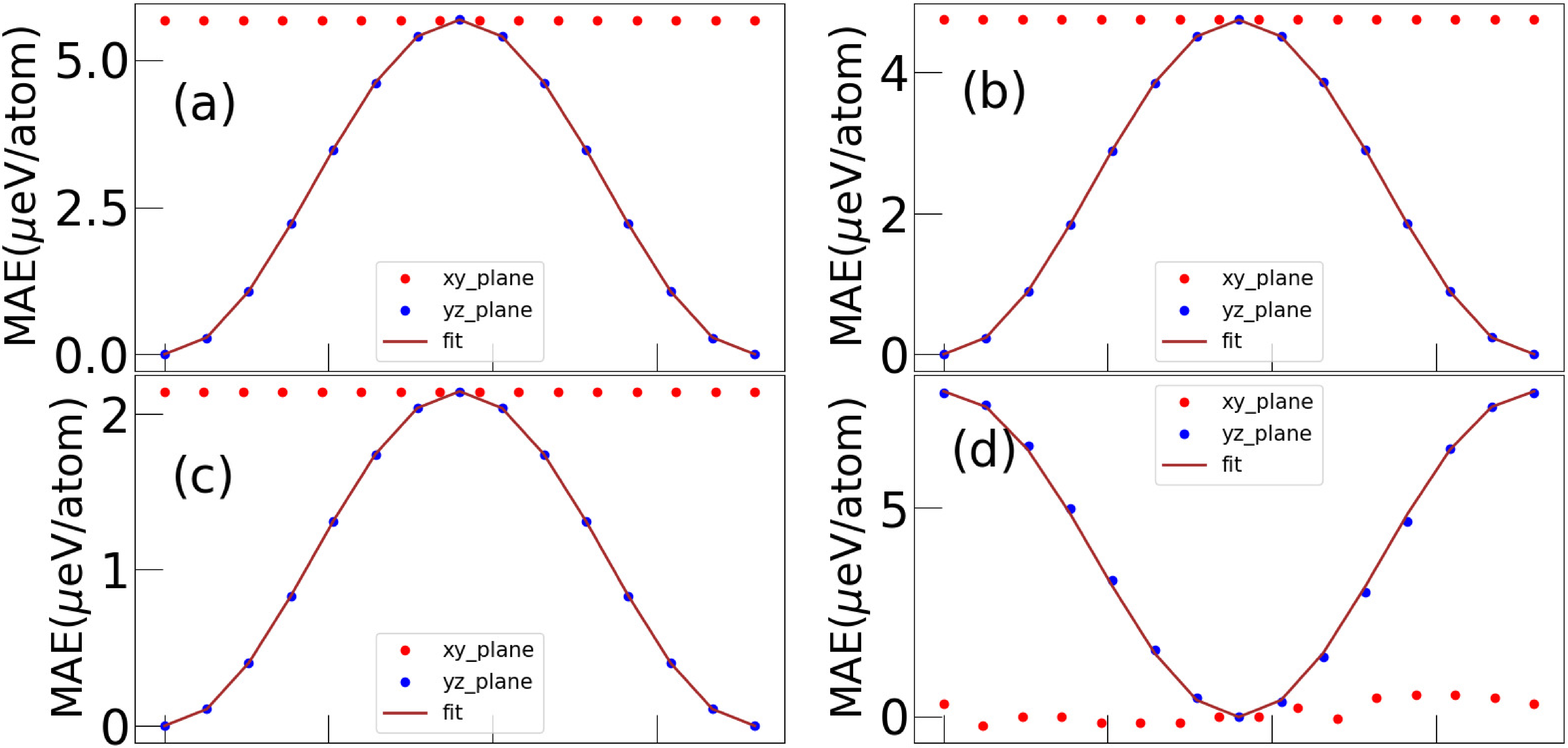}
     
   \end{subfigure}
    \hspace{0.1 mm}
   \begin{subfigure}{0.50\textwidth}
     \centering
      \includegraphics[height=6cm, width=8.20 cm]{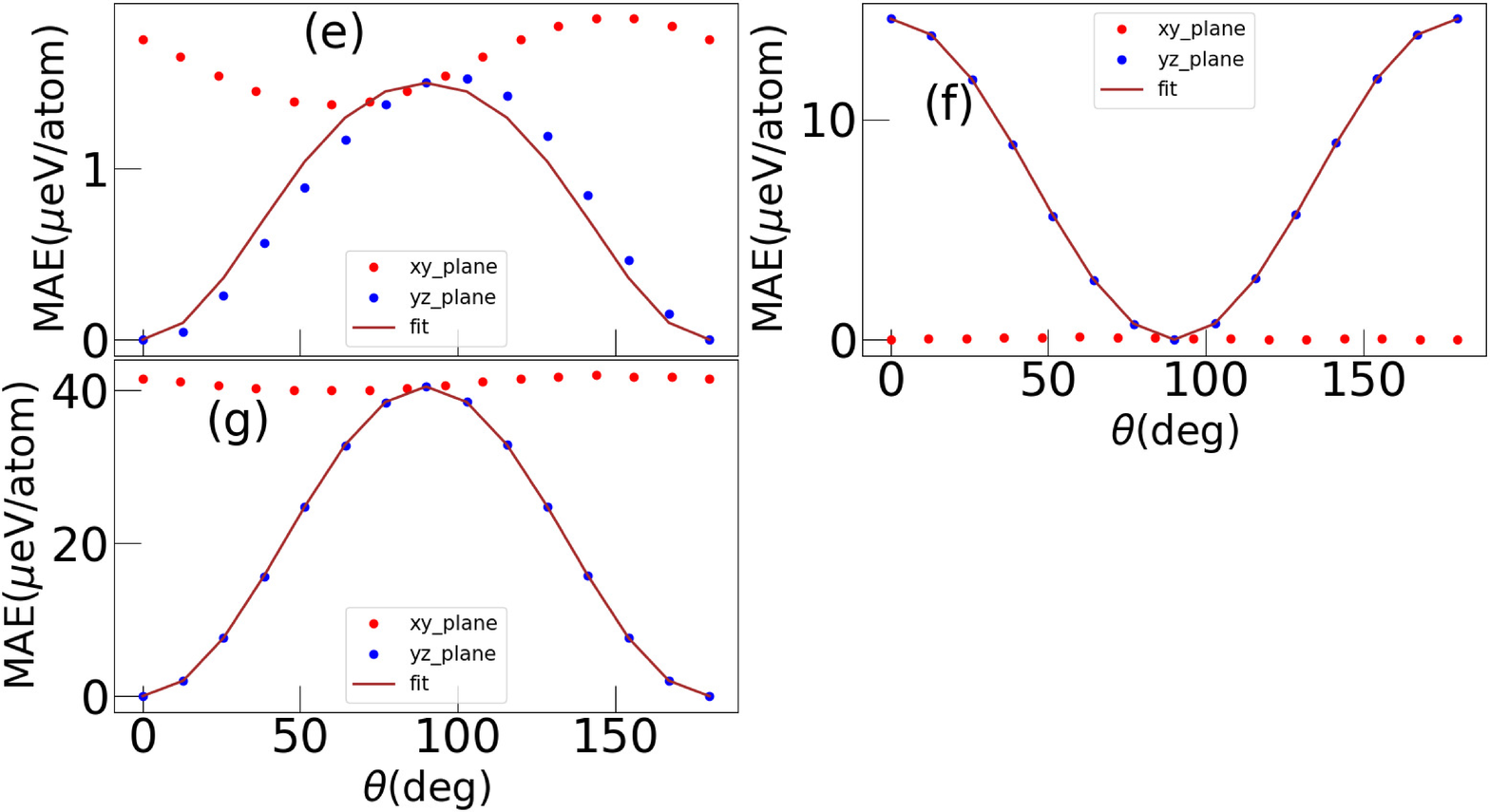}    
   \end{subfigure}
   \caption{Angular dependence of MAE in Ti\textsubscript{2}C (a) monolayer  (b)bilayer Model 1 and (c) bilayer Model 2. Results for Fe$_{2}$C in monolayer and bilayer Model 1,2,3 are shown in (d),(e),(f),(g), respectively. }
   \label{Fig:8}
 \end{figure}

 To gain more insight, the angular dependence of MAE is calculated. The results are shown in Figure ~\ref{Fig:8}. For this we calculated the MAEs of the $xy,xz$ and $yz$ planes. The MAE of $xz$ and $yz$ plane are equal for all the structures. Thus MAE of only $xy$ and $yz$ planes are shown in Figure ~\ref{Fig:8}. The spin angle $\theta$ is varied from 0 to $180^{\circ}$ at an interval of  $ 12^{\circ} $. All systems of Ti\textsubscript{2}C MXene show MAE  minima at  $ \theta = 0 ^{\circ}$ and  $ \theta = 180 ^{\circ} $.This implies  $ \theta = 0^{\circ} $ that is the $c$-axis or (001) direction is the easy direction.A similar conclusion also holds for Fe\textsubscript{2}C bilayer Model 1 and Model 3. However, for monolayer Fe\textsubscript{2}C and bilayer Model 2 , the minima of MAE are in the $xy$ plane, indicating that the $xy$ plane is the easy plane of magnetization.The dependence of MAE on spin angle $\theta$ for a hexagonal system is given by the relation\cite{siriwardane2020engineering}:

\begin{equation}
MAE (\theta)= K_{0} + K_{1}Sin^{2}\theta + K_{2}Sin^{4}\theta
\label{Eqn:5}
\end{equation}
%\lipsum
 %\lipsum
Here $K_{1}$ and $K_{2}$ are magnetic
anisotropy constants, $ \theta $ is the angle made by the spin vector with $z$ and $x$-axis for MAE in the $xy$ and $yz$ planes, respectively.If $ K_{1} > 0 $ and $ K_{1} > -K_{2} $,  the out of plane direction ($ \theta = 0 ^{\circ} $) is the direction of  magnetic easy axis. On the other hand, if $ K_{1} > 0$ and $ K_{1} < -K_{2} $ or $ K_{2} < 0 $ and $ K_{1} <-2K_{2} $, the in-plane directions($ \theta $ =  $ 90 ^{\circ} $) are the magnetic easy directions. By fitting the calculated values to the above equation, we get $K_{1} = 5.693  \mu$eV  and $K_{2} = 0.004  \mu$eV  for Ti\textsubscript{2}C monolayer. As $ K_{1} > 0 $ and $ K_{1} >-K_{2} $, we have an easy axis of magnetisation. For bilayer Ti$_{2}$C, the numbers are $K_{1}=4.73(2.14) \mu$eV, $K_{2}= 0.004(-0.001) \mu$eV for Model 1(Model 2). As $ K_{1} > 0 $ and $ K_{1} >-K_{2} $,we have  easy axes of magnetisation for both cases. In Fe\textsubscript{2}C monolayer, $K_{1}=-7.772 \mu$eV, $K_{2}=-0.267 \mu$eV. In this case, $ K_{2}<0, K_{1} < -2K_{2} $ imply an easy plane of magnetisation. In case of Fe\textsubscript{2}C bilayer, $K_{1}= 4.733(40.56) \mu$eV and $K_{2}=0.006(-0.041) \mu$eV for Model 1 (Model 3). Since $ K_{1} >0 $ and $ K_{1} >-K_{2} $ in both models, we have an easy axis of magnetisation. For  Model 2  $K_{1}= -14.562 \mu$eV, $K_{2} = -0.018 \mu$ eV imply the existence of an easy plane of magnetisation.

To understand the origin of the changes in  MAE on moving from monolayer to bilayers of Fe$_{2}$C, we use a perturbative analysis. According to second-order perturbation theory, the MAE is given as  MAE = $E_{[100]}-E_{[001]} \sim \zeta^{2}\sum_{o,u} \frac{{\vert}\langle d_{u}\vert H_{SO_{z}}\vert d_{0}\rangle\vert - \vert \langle d_{u}\vert H_{SO_{x}}\vert d_{0}\rangle\vert}{\epsilon_{u}-\epsilon_{o}} $.Here d\textsubscript{u} and d\textsubscript{o} are the unoccupied and occupied states, respectively\cite{song2019surface,ruiz2013tuning,fang2018large}.Only the  $d$ states of Fe atoms close to the fermi level  contribute majorly to the MAE values. The SOC matrix elements
$ {\vert}\langle d_{xz}\vert H_{SO}\vert d_{yz}\rangle\vert $ ,$ {\vert}\langle d_{x^{2}-y^{2}}\vert H_{SO}\vert d_{xy}\rangle\vert $ favour out of plane anisotropy whereas $ {\vert}\langle d_{x^{2}-y^{2}}\vert H_{SO}\vert d_{yz}\rangle\vert $ ,$ {\vert}\langle d_{xy}\vert H_{SO}\vert d_{xz}\rangle\vert $ and $ {\vert}\langle d_{z^{2}}\vert H_{SO}\vert d_{yz}\rangle\vert $ favour in-plane anisotropy. From the PDOS of monolayer Fe\textsubscript{2}C, we find that it is mainly the minority spin states that contribute to MAE since there is not much  contribution from the majority spin band to the states near the fermi level. In the unoccupied part of the minority band, we find a  prominent peak around 0.75 eV due to  d\textsubscript{z\textsuperscript{2}} orbitals along with other less prominent ones due to  d\textsubscript{xy}/d\textsubscript{x\textsuperscript{2}-y\textsuperscript{2}} states. In bilayer Model 1, the contribution of d\textsubscript{xy}/d\textsubscript{x\textsuperscript{2}-y\textsuperscript{2}} states in the unoccupied part increase substantially  along with a slight increase in the d\textsubscript{xz}/d\textsubscript{yz} contributions. Consequently, the contributions in the occupied part slightly decrease. This causes the enhancements in the matrix elements favouring out of plane MAE. The existence of magnetic easy axis is thus explained. In Model 2, the states contributed by  d\textsubscript{xz}/d\textsubscript{yz} and d\textsubscript{xy}/d\textsubscript{x\textsuperscript{2}-y\textsuperscript{2}} in the unoccupied part are greatly reduced as compared to Model 1.As a result contributions of the  terms to favour out of plane magnetic anisotropy are reduced. This causes further enhancement in the in-plane MAE value compared to monolayer since now both terms contributing to out-of-plane anisotropy decrease.  In Model 3, d\textsubscript{xz}/d\textsubscript{yz} and d\textsubscript{xy}/d\textsubscript{x\textsuperscript{2}-y\textsuperscript{2}} contributions increase in both the occupied and unoccupied parts due to  a significant amount of states near the fermi level in the majority spin channel. This causes enhancement of both the terms that favour out-of-plane anisotropy. Thus we get a clear picture regarding stacking dependence in the qualitative behaviour of magnetic anisotropy in Fe$_{2}$C from electronic structure.

\subsection{Conclusions}
Using first-principles calculations, we have investigated the dependence of  electronic and magnetic properties in on layer thickness and stacking patterns in Ti$_{2}$C and Fe$_{2}$C MXenes. To our knowledge, this is the first work exploring this aspect in magnetic MXenes. We find that both layer thickness and stacking patterns alter the magnetic properties in MXenes, irrespective of whether the transition metal is a weakly magnetic like Ti or a strong magnet like Fe. For Ti$_{2}$C the ground state magnetic configuration changes as one increases the number of layers; the magnetism is driven by the dangling bonds associated with the Ti atoms on the outer surface. The qualitative nature of magnetic exchange interactions and magnetic anisotropy in this system remain same in different stacking patterns. Relatively weak qualitative nature of the magnetic exchange parameters lead to magnetic transition temperatures below the room temperature. In contrast, Fe$_{2}$C is more fascinating. Here, the itinerant nature of Fe atoms contribute to the magnetic properties. Thus, unlike Ti$_{2}$C where the magnetic moments are localised on the outer surface Ti atoms, Fe atoms on both surfaces are spin polarised with the magnitudes much larger than that in Ti$_{2}$C. The stacking patterns give rise to competing superexchange and direct exchange among the Fe atoms leading to differences in the ground state of bilayer Fe$_{2}$C. This affects the qualitative nature of the magnetic exchange parameters and subsequently the magnitudes of the magnetic transition temperatures. The significant changes in the electronic structures as one changes the stacking pattern in bilayer Fe$_{2}$C lead to differences in the nature of the magnetic anisotropy in this system. Thus we conclude that there is a better scope in manipulating the magnetism in Fe$_{2}$C by varying the layer thickness and stacking pattern. This would be desirable from the point of view of device fabrication. This study opens up the possibility of further exploration into the structure-magnetism relations in magnetic MXenes.


\begin{thebibliography}{10}

\bibitem{novoselov2004electric}
Kostya~S Novoselov, Andre~K Geim, Sergei~V Morozov, De-eng Jiang, Yanshui
  Zhang, Sergey~V Dubonos, Irina~V Grigorieva, and Alexandr~A Firsov.
%\newblock Electric field effect in atomically thin carbon films.
\newblock {\em science}, 306(5696):666--669, 2004.

\bibitem{chhowalla2015two}
Manish Chhowalla, Zhongfan Liu, and Hua Zhang.
%\newblock Two-dimensional transition metal dichalcogenide (tmd) nanosheets.
\newblock {\em Chemical Society Reviews}, 44(9):2584--2586, 2015.

\bibitem{pacile2008two}
D~Pacile, JC~Meyer, {\c{C}}{\"O}~Girit, and AJAPL Zettl.
%\newblock The two-dimensional phase of boron nitride: Few-atomic-layer sheets
  and suspended membranes.
\newblock {\em Applied Physics Letters}, 92(13):133107, 2008.

\bibitem{vogt2012silicene}
Patrick Vogt, Paola De~Padova, Claudio Quaresima, Jose Avila, Emmanouil
  Frantzeskakis, Maria~Carmen Asensio, Andrea Resta, B{\'e}n{\'e}dicte Ealet,
  and Guy Le~Lay.
%\newblock Silicene: compelling experimental evidence for graphenelike
 % two-dimensional silicon.
\newblock {\em Physical review letters}, 108(15):155501, 2012.

\bibitem{kou2015phosphorene}
Liangzhi Kou, Changfeng Chen, and Sean~C Smith.
%\newblock Phosphorene: fabrication, properties, and applications.
\newblock {\em The journal of physical chemistry letters}, 6(14):2794--2805,
  2015.

\bibitem{roy2014field}
Tania Roy, Mahmut Tosun, Jeong~Seuk Kang, Angada~B Sachid, Sujay~B Desai, Mark
  Hettick, Chenming~C Hu, and Ali Javey.
%\newblock Field-effect transistors built from all two-dimensional material
 % components.
\newblock {\em ACS nano}, 8(6):6259--6264, 2014.

\bibitem{kajale20212d}
Shivam~Nitin Kajale, Shubham Yadav, Yubin Cai, Baju Joy, and Deblina Sarkar.
%\newblock 2d material based field effect transistors and nanoelectromechanical
  %systems for sensing applications.
\newblock {\em Iscience}, 24(12):103513, 2021.

\bibitem{zhang2018two}
Panpan Zhang, Faxing Wang, Minghao Yu, Xiaodong Zhuang, and Xinliang Feng.
%\newblock Two-dimensional materials for miniaturized energy storage devices:
 % from individual devices to smart integrated systems.
\newblock {\em Chemical Society Reviews}, 47(19):7426--7451, 2018.

\bibitem{subbanna20222d}
B~Bala Subbanna, Kuldeep Choudhary, Sonika Singh, and Santosh Kumar.
%\newblock 2d material-based optical sensors: a review.
\newblock {\em ISSS Journal of Micro and Smart Systems}, pages 1--9, 2022.

\bibitem{mermin}
N~D Mermin and H~Wigner.
\newblock {\em Physical Review Letters}, 17:1133, 1966.

\bibitem{huang2017layer}
Bevin Huang, Genevieve Clark, Efr{\'e}n Navarro-Moratalla, Dahlia~R Klein, Ran
  Cheng, Kyle~L Seyler, Ding Zhong, Emma Schmidgall, Michael~A McGuire, David~H
  Cobden, et~al.
%\newblock Layer-dependent ferromagnetism in a van der waals crystal down to the
  %monolayer limit.
\newblock {\em Nature}, 546(7657):270--273, 2017.

\bibitem{wang2021ferromagnetism}
Xiong Wang, Dian Li, Zejun Li, Changzheng Wu, Chi-Ming Che, Gang Chen, and
  Xiaodong Cui.
%\newblock Ferromagnetism in 2d vanadium diselenide.
\newblock {\em ACS nano}, 15(10):16236--16241, 2021.

\bibitem{gong2017discovery}
Cheng Gong, Lin Li, Zhenglu Li, Huiwen Ji, Alex Stern, Yang Xia, Ting Cao, Wei
  Bao, Chenzhe Wang, Yuan Wang, et~al.
%\newblock Discovery of intrinsic ferromagnetism in two-dimensional van der
 % waals crystals.
\newblock {\em Nature}, 546(7657):265--269, 2017.

\bibitem{fei2018two}
Zaiyao Fei, Bevin Huang, Paul Malinowski, Wenbo Wang, Tiancheng Song, Joshua
  Sanchez, Wang Yao, Di~Xiao, Xiaoyang Zhu, Andrew~F May, et~al.
%\newblock Two-dimensional itinerant ferromagnetism in atomically thin
 % fe$_3$gete$_2$.
\newblock {\em Nature materials}, 17(9):778--782, 2018.

\bibitem{naguib2011two}
Michael Naguib, Murat Kurtoglu, Volker Presser, Jun Lu, Junjie Niu, Min Heon,
  Lars Hultman, Yury Gogotsi, and Michel~W Barsoum.
%\newblock Two-dimensional nanocrystals produced by exfoliation of
%  ti$_3$alc$_2$.
\newblock {\em Advanced materials}, 23(37):4248--4253, 2011.

\bibitem{naguib2012two}
Michael Naguib, Olha Mashtalir, Joshua Carle, Volker Presser, Jun Lu, Lars
  Hultman, Yury Gogotsi, and Michel~W Barsoum.
%\newblock Two-dimensional transition metal carbides.
\newblock {\em ACS nano}, 6(2):1322--1331, 2012.

\bibitem{naguib2013new}
Michael Naguib, Joseph Halim, Jun Lu, Kevin~M Cook, Lars Hultman, Yury Gogotsi,
  and Michel~W Barsoum.
%\newblock New two-dimensional niobium and vanadium carbides as promising
 % materials for li-ion batteries.
\newblock {\em Journal of the American Chemical Society}, 135(43):15966--15969,
  2013.

\bibitem{soundiraraju2017two}
Bhuvaneswari Soundiraraju and Benny~Kattikkanal George.
%\newblock Two-dimensional titanium nitride (ti$_2$n) mxene: synthesis,
 % characterization, and potential application as surface-enhanced raman
%  scattering substrate.
\newblock {\em ACS nano}, 11(9):8892--8900, 2017.

\bibitem{meshkian2015synthesis}
Rahele Meshkian, Lars-{\AA}ke N{\"a}slund, Joseph Halim, Jun Lu, Michel~W
  Barsoum, and Johanna Rosen.
%\newblock Synthesis of two-dimensional molybdenum carbide, mo$_2$c, from the
 % gallium based atomic laminate mo$_2$ga$_2$c.
\newblock {\em Scripta Materialia}, 108:147--150, 2015.

\bibitem{urbankowski20172d}
Patrick Urbankowski, Babak Anasori, Kanit Hantanasirisakul, Long Yang, Lihua
  Zhang, Bernard Haines, Steven~J May, Simon~JL Billinge, and Yury Gogotsi.
%\newblock 2d molybdenum and vanadium nitrides synthesized by ammoniation of 2d
 % transition metal carbides (mxenes).
\newblock {\em Nanoscale}, 9(45):17722--17730, 2017.

\bibitem{urbankowski2016synthesis}
Patrick Urbankowski, Babak Anasori, Taron Makaryan, Dequan Er, Sankalp Kota,
  Patrick~L Walsh, Mengqiang Zhao, Vivek~B Shenoy, Michel~W Barsoum, and Yury
  Gogotsi.
%\newblock Synthesis of two-dimensional titanium nitride ti$_4$n$_3$ (mxene).
\newblock {\em Nanoscale}, 8(22):11385--11391, 2016.

\bibitem{zhou2016two}
Jie Zhou, Xianhu Zha, Fan~Y Chen, Qun Ye, Per Eklund, Shiyu Du, and Qing Huang.
%\newblock A two-dimensional zirconium carbide by selective etching of
  %al$_3$c$_3$ from nanolaminated zr$_3$al$_3$c$_5$.
\newblock {\em Angewandte Chemie International Edition}, 55(16):5008--5013,
  2016.

\bibitem{ma2021ti}
Rui Ma, Zetong Chen, Danna Zhao, Xujing Zhang, Jingting Zhuo, Yajiang Yin,
  Xiaofeng Wang, Guowei Yang, and Fang Yi.
%\newblock Ti$_3$c$_2$t$_x$ mxene for electrode materials of supercapacitors.
\newblock {\em Journal of Materials Chemistry A}, 9(19):11501--11529, 2021.

\bibitem{zheng2022mxene}
Wei Zheng, Joseph Halim, Johanna Rosen, Michel~W Barsoum, et~al.
%\newblock Mxene-based symmetric supercapacitors with high voltage and high
 % energy density.
\newblock {\em Materials Reports: Energy}, 2(1):100078, 2022.

\bibitem{si2015half}
Chen Si, Jian Zhou, and Zhimei Sun.
%\newblock Half-metallic ferromagnetism and surface functionalization-induced
  %metal--insulator transition in graphene-like two-dimensional cr$_2$c
  crystals.
\newblock {\em ACS applied materials \& interfaces}, 7(31):17510--17515, 2015.

\bibitem{sun2021cr2nx2}
Qian Sun, Jie Li, Yi~Li, Zongxian Yang, and Ruqian Wu.
%\newblock Cr$_2$nx$_2$ mxene (x= o, f, oh): A 2d ferromagnetic half-metal.
\newblock {\em Applied Physics Letters}, 119(6):062404, 2021.

\bibitem{yue2021tuning}
Yunliang Yue, Buwei Wang, Nanxi Miao, Chao Jiang, Hongwei Lu, Bowen Zhang,
  Yankai Wu, Jie Ren, and Min Wang.
%\newblock Tuning the magnetic properties of zr$_2$n mxene by biaxial strain.
\newblock {\em Ceramics International}, 47(2):2367--2373, 2021.

\bibitem{he2016new}
Junjie He, Pengbo Lyu, and Petr Nachtigall.
%\newblock New two-dimensional mn-based mxenes with room-temperature
 % ferromagnetism and half-metallicity.
\newblock {\em Journal of Materials Chemistry C}, 4(47):11143--11149, 2016.

\bibitem{zhang2022computational}
Yinggan Zhang, Zhou Cui, Baisheng Sa, Naihua Miao, Jian Zhou, and Zhimei Sun.
%\newblock Computational design of double transition metal mxenes with intrinsic
 % magnetic properties.
\newblock {\em Nanoscale Horizons}, 7(3):276--287, 2022.

\bibitem{sivadas2018stacking}
Nikhil Sivadas, Satoshi Okamoto, Xiaodong Xu, Craig~J Fennie, and Di~Xiao.
%\newblock Stacking-dependent magnetism in bilayer cri$_3$.
\newblock {\em Nano letters}, 18(12):7658--7664, 2018.

\bibitem{science2019}
Weijong Chen, Zeyuan Sun, Zhongjie Wang, Lehua Gu, Xiaodong Xu, Shiwei Wu, and
  Chunlei Gao.
\newblock {\em Science}, 336:983, 2019.

\bibitem{mxene-multilayer2019}
Cui Jin, Jing Shang, Tang Xiao, Xin Tan, Sean~C Smith, Chengwang Niu, Ying Dai,
  and Liangzhi Kou.
\newblock {\em Journal of Materials Chemistry C}, 7:15308, 2019.

\bibitem{kong2021switching}
Xiangru Kong, Hongkee Yoon, Myung~Joon Han, and Liangbo Liang.
%\newblock Switching interlayer magnetic order in bilayer cri$_3$ by stacking
 % reversal.
\newblock {\em Nanoscale}, 13(38):16172--16181, 2021.

\bibitem{wang2021systematic}
Duo Wang and Biplab Sanyal.
%\newblock Systematic study of monolayer to trilayer cri$_3$: Stacking sequence
  %dependence of electronic structure and magnetism.
\newblock {\em The Journal of Physical Chemistry C}, 125(33):18467--18473,
  2021.

\bibitem{lv2020monolayer}
Peng Lv, Yan-Li Li, and Jia-Fu Wang.
%\newblock Monolayer ti $_2$c mxene: Manipulating magnetic properties and
 % electronic structures by an electric field.
\newblock {\em Physical Chemistry Chemical Physics}, 22(20):11266--11272, 2020.

\bibitem{yue2017fe2c}
Yunliang Yue.
%\newblock Fe$_2$c monolayer: An intrinsic ferromagnetic mxene.
\newblock {\em Journal of Magnetism and Magnetic Materials}, 434:164--168,
  2017.

\bibitem{dft}
W~Kohn and L~J Sham.
\newblock {\em Physical Review}, 140:A1133, 1965.

\bibitem{kresse1999ultrasoft}
Georg Kresse and Daniel Joubert.
%\newblock From ultrasoft pseudopotentials to the projector augmented-wave
 % method.
\newblock {\em Physical review b}, 59(3):1758, 1999.

\bibitem{kresse1996efficient}
Georg Kresse and J{\"u}rgen Furthm{\"u}ller.
%\newblock Efficient iterative schemes for ab initio total-energy calculations
  %using a plane-wave basis set.
\newblock {\em Physical review B}, 54(16):11169, 1996.

\bibitem{perdew1996generalized}
John~P Perdew, Kieron Burke, and Matthias Ernzerhof.
%\newblock Generalized gradient approximation made simple.
\newblock {\em Physical review letters}, 77(18):3865, 1996.

\bibitem{grimme2010consistent}
Stefan Grimme, Jens Antony, Stephan Ehrlich, and Helge Krieg.
%\newblock A consistent and accurate ab initio parametrization of density
%  functional dispersion correction (dft-d) for the 94 elements h-pu.
\newblock {\em The Journal of chemical physics}, 132(15):154104, 2010.

\bibitem{monkhorst1976special}
Hendrik~J Monkhorst and James~D Pack.
%\newblock Special points for brillouin-zone integrations.
\newblock {\em Physical review B}, 13(12):5188, 1976.

\bibitem{eriksson2017atomistic}
Olle Eriksson, Anders Bergman, Lars Bergqvist, and Johan Hellsvik.
%\newblock {\em Atomistic spin dynamics: foundations and applications}.
\newblock Oxford university press, 2017.

\bibitem{huang2021theoretical}
Haiming Huang, Weiliang Wang, and Shaolin Zhang.
%\newblock Theoretical assessment of raman spectra on mxene ti $_2$c: from
 % monolayer to bilayer.
\newblock {\em Physical Chemistry Chemical Physics}, 23(35):19884--19891, 2021.

\bibitem{GaN}
X~G Zhao, Z~Tang, and W~X Hu.
\newblock {\em Surface Science}, 608:97, 2013.

\bibitem{luo2020first}
Kan Luo, Xian-Hu Zha, Qing Huang, Cheng-Te Lin, Minghui Yang, Shenghu Zhou, and
  Shiyu Du.
%\newblock First-principles study of magnetism in some novel mxene materials.
\newblock {\em RSC advances}, 10(72):44430--44436, 2020.

\bibitem{stohr2006magnetism}
Joachim St{\"o}hr and Hans~Christoph Siegmann.
\newblock Magnetism solid-state sciences 152, 2006.

\bibitem{siriwardane2020engineering}
Edirisuriya~MD Siriwardane, Pragalv Karki, Yen~Lee Loh, and Deniz
  {\c{C}}ak{\i}r.
%\newblock Engineering magnetic anisotropy and exchange couplings in double
%  transition metal mxenes via surface defects.
\newblock {\em Journal of Physics: Condensed Matter}, 33(3):035801, 2020.

\bibitem{song2019surface}
Changsheng Song, Xin Liu, Xiaoping Wu, Jingjing Wang, Jiaqi Pan, Tingyu Zhao,
  Chaorong Li, and Jiqing Wang.
%\newblock Surface-vacancy-induced metallicity and layer-dependent magnetic
 % anisotropy energy in cr$_2$ge$_2$te$_6$.
\newblock {\em Journal of Applied Physics}, 126(10):105111, 2019.

\bibitem{ruiz2013tuning}
P~Ruiz-D{\'\i}az, TR~Dasa, and VS~Stepanyuk.
%\newblock Tuning magnetic anisotropy in metallic multilayers by surface
 % charging: an ab initio study.
\newblock {\em Physical review letters}, 110(26):267203, 2013.

\bibitem{fang2018large}
Yimei Fang, Shunqing Wu, Zi-Zhong Zhu, and Guang-Yu Guo.
%\newblock Large magneto-optical effects and magnetic anisotropy energy in
 % two-dimensional cr$_2$ge$_2$te$_6$.
\newblock {\em Physical Review B}, 98(12):125416, 2018.

\end{thebibliography}
\end{document}